\documentclass[11pt,a4paper]{article}
\usepackage{graphicx}
\usepackage{jcappub}

\title{Enhanced Sensitivity to Dark Matter Self-annihilations in the Sun using Neutrino Spectral Information}

\author[a]{C.~Rott}
\author[b]{T.~Tanaka}
\author[b,c]{Y.~Itow}

\affiliation[a]{Dept.~of Physics and Center for Cosmology and Astro-Particle Physics,\\
Ohio State University, Columbus, OH 43210, USA}
\affiliation[b]{Solar-Terrestrial Environment Laboratory, Nagoya University,\\
Furo-cho, Chikusa-ku, Nagoya, 464-8601, Japan}
\affiliation[c]{Kobayashi-Maskawa Institute for the Origin of Particles and the Universe, \\
Nagoya University, Furo-cho, Chikusa-ku, Nagoya, 464-8601, Japan}

\emailAdd{carott@mps.ohio-state.edu}
\emailAdd{ttanaka@stelab.nagoya-u.ac.jp}
\emailAdd{itow@stelab.nagoya-u.ac.jp}

\abstract{Self-annihilating dark matter gravitationally captured by the Sun could yield observable neutrino signals at current and next generation neutrino detectors.
By exploiting such signals, neutrino detectors can probe the spin-dependent scattering of weakly interacting massive particles (WIMPs) with nucleons in the Sun. 
We describe a method how to convert constraints on neutrino fluxes to a limit on the WIMP-nucleon scattering cross section. In this method all neutrino flavors can be treated in a very similar way.
We study the sensitivity of neutrino telescopes for Solar WIMP signals using vertex contained events and find that this detection channel is of particular importance in the search for low mass WIMPs. We obtain highly competitive sensitivities with all neutrino flavor channels for a Megaton sized detector through the application of basic spectral selection criteria.  
Best results are obtained with the electron neutrino channel.
We discuss associated uncertainties and provide a procedure how to treat them for analyses in a consistent way.}

\notoc

\keywords{Dark Matter, Solar WIMPs, WIMP-nucleon scattering cross section, Neutrino Astronomy}

\begin{document}

\maketitle
\flushbottom



\section{Introduction}

There is compelling observational evidence for the existence of dark matter; however, its nature remains unknown. 
A prominent candidate that naturally arises in supersymmetric~\cite{Martin:1997ns} and large extra dimension models~\cite{Appelquist:2000nn} is a weakly interacting massive particle (WIMP)~\cite{Steigman:1984ac}. WIMPs are predicted to have masses ranging from a few GeV to several TeV~\cite{Bertone:2004pz}. They could be detected through a variety of possible signals, indirectly through the byproducts of self-annihilations or decays, directly by scattering effects with nuclei, or production at colliders.

The nucleon coupling of a WIMP in the extreme nonrelativistic limit is characterized by a scalar (spin-independent) and an axial vector (spin-dependent) term.  
In general the spin-dependent (SD) couplings for WIMPs on protons and neutrons are expected to be of the same order of magnitude, and very similar for the spin-independent (SI) case~\cite{PhysRevD.40.3132}. Throughout the paper we assume that the WIMP-proton and WIMP-neutron scattering cross sections are identical and do not consider isospin violating interactions, which have been discussed elsewhere (see for example ref.~\cite{Feng:2011vu}).

The spin-dependent and spin-independent terms behave very differently if coherence across the nucleus is taken into account; the nucleon contributions interfere constructively to enhance the WIMP-nucleus elastic cross section. The WIMP-nucleon cross sections for spin-independent interactions are generally orders of magnitude smaller than those allowed for the spin-dependent case. Constructive interference across the nucleus leads to a quadratic enhancement with the nucleon number $(\sim A^2)$ in the spin-independent couplings. By exploiting this dependence through the use of high $Z$ elements, direct detection experiments such as CDMS-II~\cite{Ahmed:2009zw}, XENON100~\cite{Aprile:2011hi}, and EDELWEISS-II~\cite{Armengaud:2011cy} have been able to tightly constrain the spin-independent WIMP-nucleon scattering cross section.

Limits on the spin-independent cross section extend to $10^{-7}~{\rm pb}$ ($10^{-43}~{\rm cm}^2$) for WIMPs mass around 50~GeV. For the spin-dependent case constraints from direct detection experiments, such as SIMPLE~\cite{Felizardo:2010mi}, COUPP~\cite{Behnke:2010xt}, PICASSO~\cite{Archambault:2009sm} are much weaker and range down to a minimum of about $2.8 \times 10^{-2}$~pb for a WIMP mass of 45~GeV. 
Better sensitivities are achieved by indirect searches with neutrino telescopes for this case. Depending on the assumed annihilation channel, limits for the spin-dependent case from Super-Kamiokande~\cite{Desai:2004pq} and IceCube~\cite{Abbasi:2009uz} range from $10^{-2}$~pb to $10^{-4}$~pb. 
These tight constraints are obtained by looking for neutrino signals from self-annihilating dark matter gravitationally captured by the Sun. The comparison becomes possible, as the capture process is initiated by the same interaction that direct detection experiments exploit: a nuclear recoil in which a WIMP transfers a fraction of its kinetic energy to a nucleon. For the commonly assumed case of equilibrium between capture rate, $\Gamma_{\rm C}$, and annihilation rate, $\Gamma_{\rm A}$, neutrino fluxes, $\phi_{\nu}^{f}$, from the Sun are directly dependent on the WIMP-nucleon scattering cross section. 
Hence, the neutrino flux from the Sun or any constraint on it can directly be linked to the WIMP-nucleon scattering cross section.

While previous work has focused on the problem of converting the WIMP-nucleon scattering cross sections to muon neutrino generated muon 
flux, $\phi_{\mu}^{f}$~\cite{Kamionkowski:1994dp,Wikstrom:2009kw}, we here focus on the neutrino flux instead.
Muon fluxes inherently contain a dependence on the detector location as well as assumptions on neutrino interactions and muon propagation through some medium.
Using the neutrino flux itself has the benefit that it allows to treat all neutrino flavors in a very similar way and the muon neutrino induced muon flux is just one special case of it. The neutrino flux is directly related to the study of vertex contained events at neutrino detectors. This class of events is in particular relevant for precision measurements of neutrino flavor ratios and the extraction of neutrino spectral information. We discuss results in light of a generic neutrino detector. Results are applicable for example to Super-Kamiokande, IceCube/DeepCore, KAMLAND, and future detectors proposed for long baseline neutrino oscillation experiments such as Hyper-K~\cite{Nakamura:2003hk} and LENA~\cite{Wurm:2011zn}. For a detector size (diameter) that is on the same order or larger compared to the average length of a muon track created by a muon neutrino with energy of interest, vertex contained events will be the dominant part of the expected signals in the detector.

This paper is structured as follows: In section~\ref{sec:FluxConversion} we provide a description how to compare a neutrino flux from the Sun to the spin-dependent WIMP-nucleon scattering cross section, $\sigma^{SD}$. Our result is based on the DarkSUSY package version 5.0.4~\cite{Gondolo:2004sc}, which allows one to obtain conversion factors for given halo models and supersymmetric model parameters. Our study is focused on the WIMP mass range favored by supersymmetry (10~GeV to 10~TeV), however results are completely model-independent and applicable to any model that produces WIMPs in this mass range. 
We discuss the prospects of observing WIMP signals at neutrino detectors in section~\ref{sec:Results}. We use a general approach that utilizes vertex contained events of all active neutrino flavors. Results are quoted for a data size of 200~kton$\cdot$years and 5~Mton$\cdot$years. 
We calculate differential neutrino fluxes, compared to detector backgrounds from atmospheric neutrinos and give event spectra. 
In section~\ref{sec:Uncertainty} we describe how to incorporate and treat uncertainties related to the flux conversion. In particular we address uncertainties due to the dark matter distribution, which is underlying to all searches. We study the equilibrium condition and discuss procedures for scenarios when it is not satisfied. 
We summarize and draw our conclusions in section~\ref{sec:Conclusions}.


\section{Neutrino Flux Conversion}
\label{sec:FluxConversion}

In this section we review the neutrino flux conversion to the WIMP-nucleon scattering cross section as implemented in DarkSUSY~\cite{Gondolo:2004sc}. We introduce the parameters and describe the choice of values used in this study.

\subsection{Review of Standard Calculation}

WIMPs from the Galactic dark matter halo could be gravitationally captured by the Sun. This process is initiated by an initial scatter of a WIMP with a nucleon in the Sun, in which the WIMP could lose enough energy to be gravitationally bound and eventually captured by the Sun. 
The number of WIMPs in the Sun, $\mathcal{N}$, is governed by the following differential equation with three constants,

\begin{equation}
 \frac{d\mathcal{N}}{dt} = C_{\rm C} - C_{\rm A} \mathcal{N}^2 -C_{\rm E} \mathcal{N}.
\end{equation}
$C_{\rm C}$ is a WIMP capture term, that depends on the WIMP-nucleon scattering cross section. It can be identified directly with the capture rate ($C_{\rm C} = \Gamma_{\rm C}$).
$C_{\rm A}$ controls annihilation and depends on the dark matter self-annihilation cross section, $\langle \sigma_{\rm A} v \rangle$.
$C_{\rm E}$ is an evaporation term, which is only relevant for WIMP masses below 4~GeV~\cite{Gaisser:1986ha,Griest:1986yu}, and is therefore outside the mass
range considered here. Neglecting $C_{\rm E}$, the annihilation rate, $\Gamma_{\rm A}$, is given by $\Gamma_{\rm A} = \frac{1}{2} C_{\rm A} \mathcal{N}^2$. As the Sun accumulates dark matter the annihilation rate at time $t$ can be linked to the capture rate via the annihilation equation. Under the assumption of a static capture rate

\begin{equation}
  \Gamma_{\rm A} = \frac{1}{2} C_{\rm C} \tanh^2 (t/\tau),
\end{equation}
with the equilibrium time scale, $\tau$, given by

\begin{equation}
  \tau = 1/\sqrt{C_{\rm C} C_{\rm A}}.
\label{label:tau}
\end{equation}
For the case that $t \gg \tau $ equilibrium between capture and annihilation is reached. 
At present, using the age of the Sun with $t = t_\odot \sim 4.7 \times 10^9$ years the equilibrium condition 
is in general fulfilled.\footnote{We discuss a treatment for deviations from the equilibrium between capture and annihilation in section~\ref{impact_of_non_equilibrium}.}
For this case the annihilation rate is regulated by the capture rate, $\Gamma_{\rm C}$ and hence independent from the dark matter self-annihilation cross section:

\begin{equation}
  \Gamma_{\rm A} = \frac{1}{2} C_{\rm C}.
\end{equation}
We follow earlier approaches~\cite{Gould:1987ir,Wikstrom:2009kw} as
implemented in DarkSUSY for the calculation of the capture rate.
For the velocity of the WIMPs outside the potential well of the Sun in the Galactic frame, $u$, we assume a Maxwellian distribution,

\begin{equation}
 f(u)/u = \sqrt {\frac{3}{2 \pi}} \frac{n_{\chi}}{v_d  v_{\odot}} \left( \exp \left(- \frac{3(u - v_{\odot})^2}{2v_d^2} \right) - \exp \left( \frac{3(u + v_{\odot})^2}{2v_d^2} \right) \right),
\label{Maxwell_vel}
\end{equation}
where $v_\odot$ is the velocity of the Sun relative to the halo, $v_d$ is the WIMP
velocity dispersion, and $n_{\chi}$ is the local WIMP number density given by the ratio of local density, $\rho_{0}$, and WIMP mass, $m_{\chi}$. 
Using this assumption on the velocity, the total capture rate can be written as,

\begin{equation}
  C_{\rm C} = \int_{0}^{R_{\odot}} 4 \pi r^2 dr \sum_i \frac{dC_i (r)}{dV},
\label{label:C_C}
\end{equation}
where $R_{\odot}$ is the radius of the Sun, $C_i$ is the WIMP capture
rate per unit shell volume of given composition from element ``$i$'' in the Sun.
We use the solar composition model BS2005-AGS~\cite{Bahcall:2004pz}.
Since the Sun is predominately a Hydrogen target WIMP-nucleon scatterings are expected to be dominated by spin-dependent interactions. 
The capture total rate $\Gamma_{\rm C}$ is the sum of a spin-independent $\Gamma_{\rm C}^{\rm SI} $ and spin-dependent $\Gamma_{\rm C}^{\rm SD} $ term. 
We assume $\Gamma_{\rm C}^{\rm SD} \gg \Gamma_{\rm C}^{\rm SI}$ and neglect spin-independent interactions in the Sun by setting the spin-independent WIMP-proton cross section to zero, so that $\Gamma_{\rm C} \approx \Gamma_{\rm C}^{\rm SD}$.
 This choice is well motivated by the very stringent constrains on $\sigma^{\rm SI}$ from direct detection experiments~\cite{Aprile:2011hi,Armengaud:2011cy,Ahmed:2009zw}. Alternatively, one can use a neutrino flux limit from the Sun to probe the spin-independent cross section by assuming $\Gamma_{\rm C} \approx \Gamma_{\rm C}^{\rm SI}$. For this condition in practice results are roughly two orders of magnitude smaller compared to the spin-dependent scattering cross section (see e.g. ref.~\cite{Kappl:2011kz}).

The WIMP capture rate per unit shell volume from element $i$ in the Sun is given~\cite{Gould:1987ir} by

\begin{equation}
\frac{dC_{i}}{dV} = \int_{0}^{u_{max}} du \frac{f(u)}{u} w \Omega_{v,i}(w),
\label{label:C_i_dV}
\end{equation}
where at the interaction point in the Sun, the WIMP has been accelerated by the corresponding escape velocity, $v$, so that its velocity is given by $w = \sqrt{u^{2} + v^{2}}$. $\Omega_{v,i}(w)$ is the capture probability per unit time for the corresponding element $i$ (we refer the reader to Wikstr{\"o}m and Edsj{\"o}~\cite{Wikstrom:2009kw}). The upper integration bound $u_{max}$ is the velocity at which the WIMP scatter to the escape velocity~\cite{Wikstrom:2009kw}.
The rough dependence of the capture rate on various quantities can be understood from the following analytical expression~\cite{Gould:1992,Mena:2007ty}:
\begin{equation}
C_{\rm C}\simeq 9 \times 10^{24} {\rm s}^{-1} \left( \frac{\rho_{0}}{0.3 {\rm GeV}/{\rm cm}^3} \right) 
\left( \frac{270 {\rm km/s}}{\bar{v}_{d}} \right)^3
\times
\left( \frac{\sigma}{10^{-2}{\rm pb}} \right) 
\left( \frac{50~{\rm GeV}}{m_{\chi}} \right)^{2}. 
\label{eq:Gould_approximation}
\end{equation}
While equation~(\ref{eq:Gould_approximation}) is good to demonstrate the dependencies, rates deviate significantly compared to the full computation with DarkSUSY, as shown in table~\ref{equi_table}, and should not be used for event rate estimates.

\begin{table*}
\caption{Capture rate as function of WIMP mass and scattering cross section $\sigma^{\rm SD}$ as obtained with DarkSUSY and compared with results from the analytic approximation eq.~(\ref{eq:Gould_approximation})~\cite{Mena:2007ty,Gould:1992}. We show the impact of including and neglecting planetary effects following the default DarkSUSY implementation of them. 
\label{equi_table}}
\begin{tabular}{|l||r|r|r|}
\hline 
WIMP mass  & \multicolumn{3}{|c|}{Capture Rate $\Gamma_{\rm C}$ / $\sigma^{SD}$ [($1/{\rm s}$)/pb]} \\
$m_{\chi}$             & \multicolumn{2}{|c|}{Jupiter} & Approximation\\ 
$({\rm GeV})$             & neglected & included & eq.~(\ref{eq:Gould_approximation})~\cite{Mena:2007ty,Gould:1992} \\ \hline
10     &  $9.43 \times 10^{27}$  &  $9.42 \times 10^{27}$ & $2.25 \times 10^{28}$ \\
50     &  $8.74 \times 10^{26}$  &  $8.69 \times 10^{26}$ & $9.00 \times 10^{26}$ \\
100    &  $2.48 \times 10^{26}$  &  $2.45 \times 10^{26}$ & $2.25 \times 10^{26}$ \\
500    &  $1.09 \times 10^{25}$  &  $1.01 \times 10^{25}$ & $0.90 \times 10^{25}$\\
1000   &  $2.75 \times 10^{24}$  &  $2.38 \times 10^{24}$ & $2.25 \times 10^{24}$\\
5000   &  $1.11 \times 10^{23}$  &  $4.82 \times 10^{22}$ & $9.00 \times 10^{22}$\\
10000  &  $2.77 \times 10^{22}$  &  $3.47 \times 10^{21}$ & $2.25 \times 10^{22}$\\
\hline
\end{tabular}
\end{table*}

The differential neutrino ($\nu_{j} + \bar{\nu}_{j}$) energy spectrum of flavor $j$ at Earth per annihilation is given by:

\begin{equation}
\label{eqn:diff_flux_at_detector}
n_{\nu_{j}}^{f} = \frac{dN_{\nu_{j}}}{dE_{\nu_{j}}} =
\sum_{i} P(j,i) \sum_{f} B_f \frac{dN_i^f}{dE_{\nu_{j}}},
\end{equation}
where $P(j,i)$ is the probability that produced
neutrino of flavor $i$ oscillates to flavor $j$ when they propagate
to the detector, $B_f$ is the branching ratio for annihilation channel
$f$, and $dN_i^f / dE_{\nu_{j}}$ the corresponding differential number of neutrinos
of flavor $i$ per annihilation.
The integrated neutrino flux for an annihilation rate, $\Gamma_{\rm A}$, above an energy threshold $E^{thr}$ is then given by
\begin{equation}
\label{eqn:flux_at_detector}
\Phi_{\nu_{j}}^{f} = 
\frac{dN^{E \ge E^{thr}}_{\nu_{j}}}{dA dt} = 
 \frac{\Gamma_{\rm A}}{4\pi D_\odot ^2}
\int_{E^{thr}}^{\infty} dE_{\nu_{j}} n_{\nu_{j}}^{f}.
\end{equation}
Here $D_{\odot}$ is the distance of the center of the Sun to Earth.

For our study we consider several benchmark annihilation channels that represent soft ($b\bar{b}$) and hard ($W^{+}W^{-}$,$\tau^{+}\tau^{-}$) neutrino spectra.
We also take into account the suppression of the capture rate due to the
gravitational effects from planets, especially Jupiter, during the
capture mechanism~\cite{Peter:2009mk,Peter:2008sy}. We use the default DarkSUSY implementation of this process.
We scan through MSSM-7 models~\cite{Bergstrom:2008gr} using DarkSUSY and enforce the equilibrium condition, so that the flux is independent of the evolution of the dark matter in the Sun. We calculate conversion factors and the differential neutrino flux from WIMP annihilations based on the assumptions and equations above.

At production the neutrino flavor ratio is given by~\cite{Cirelli:2005gh}:
\begin{equation}
\nu_{\rm e}:\bar{\nu}_{\rm e}:\nu_{\mu}:\bar{\nu}_{\mu}:\nu_{\tau}:\bar{\nu}_{\tau} = 1:1:1:1:r:r,
\end{equation}
where the factor, $r$, depends on the annihilation channel. 
Oscillation and propagation effects distort this ratio, so that at Earth the yields for muon and tau neutrinos are more or less equal. All oscillation and propagation effects were taken into account as part of DarkSUSY~\cite{Gondolo:2004sc}.
For the neutrino oscillations, we used the parameters described in
ref.~\cite{Maltoni:2004ei} as $\theta_{12} = 33.2^{\circ}$ , $\theta_{13} =
0^{\circ}$, $\theta_{23} = 45^{\circ}$, $\delta =0$, $\Delta m^2_{21} =
8.1 \times 10^{-5} {\rm eV}^2$ and $|\Delta m^2_{31}| = 2.2 \times 10^{-3} {\rm eV}^2 $. 
For a neutrino energy threshold of $E_{\nu}^{\rm thr} = 1$~GeV, we compute conversion factors for all three neutrino flavors and various annihilation channels. We assume that the local WIMP density $\rho_{0}$ is $0.3$~GeV/cm$^3$, a Maxwellian velocity distribution with a velocity dispersion of $v_{d}=270$~km/s.

\begin{figure*}
\includegraphics[width=.48\textwidth,angle=-90]{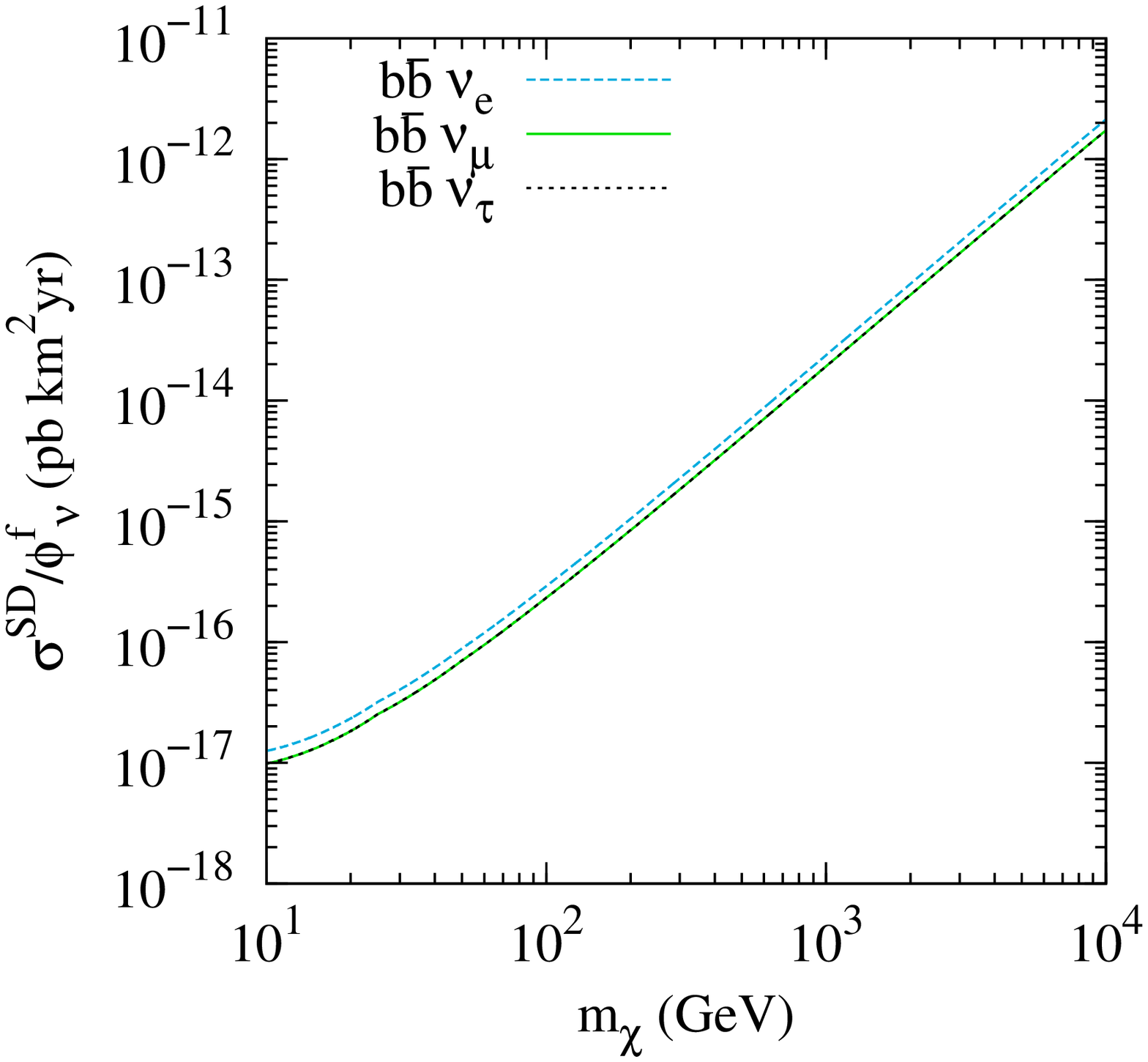}
\includegraphics[width=.48\textwidth,angle=-90]{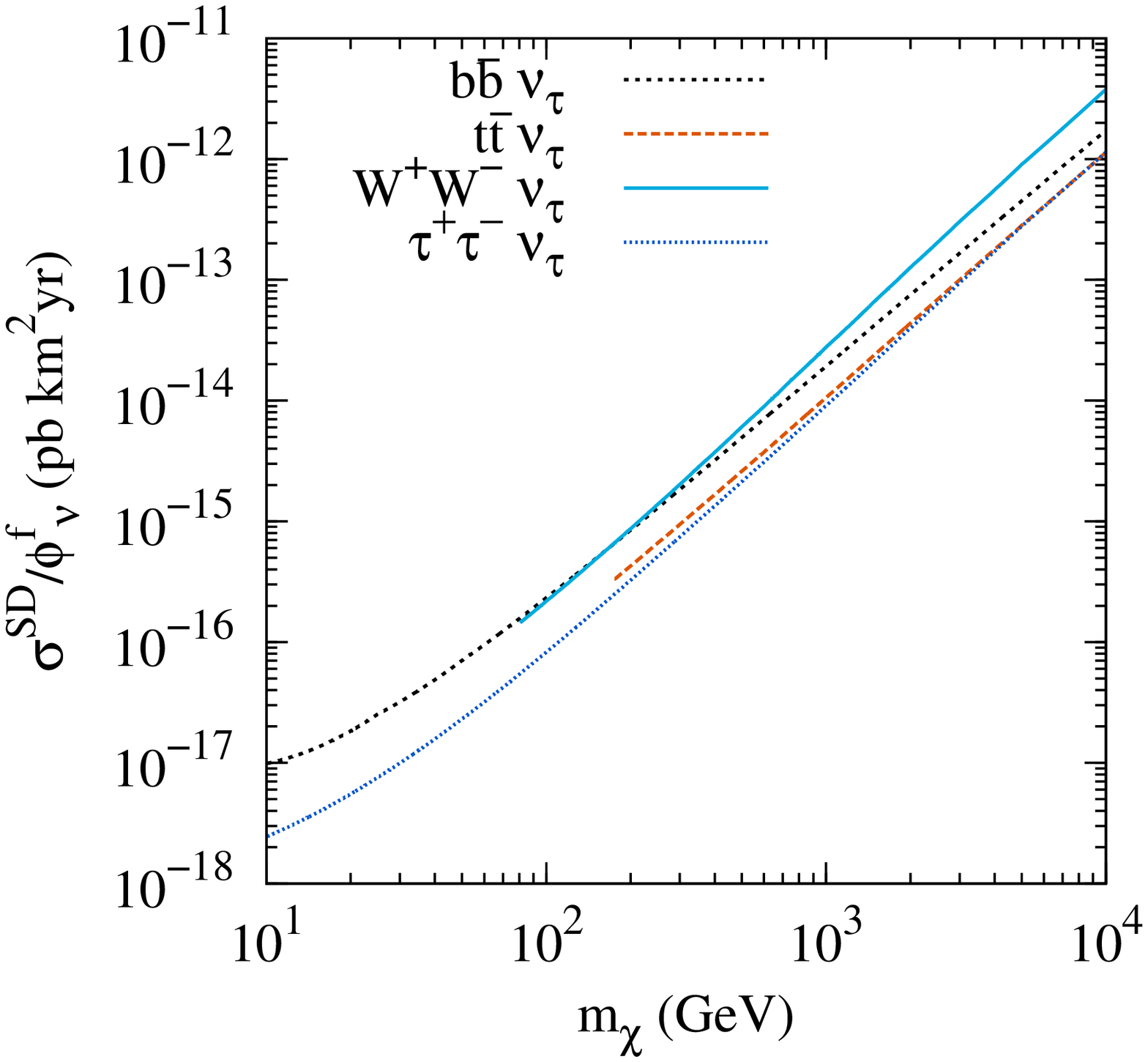}
\caption{Conversion factor, which we computed with DarkSUSY, for the integrated neutrino flux at Earth above a threshold of 1~GeV to the spin-dependent WIMP-nucleon scattering cross section obtained through DarkSUSY. Left: Neutrino flux conversion for different neutrino flavors to $\sigma^{\rm SD}$ for a given annihilation channel ($b\bar{b}$). Right: Flux conversion for various annihilation channels ($b\bar{b}, t\bar{t}, W^{+}W^{-}, \tau^{+}\tau^{-}$) for tau neutrino fluxes.} 
\label{fig1}
\end{figure*}

Figure~\ref{fig1} illustrates conversion factors for various neutrino flavors and annihilation channels that we computed with DarkSUSY. $\sigma^{\rm SD}/\phi^{f}_{\nu}$ is nearly identical for $\nu_{\mu}$ and $\nu_{\tau}$ and the difference relative to $\nu_{\rm e}$ is roughly $20$\%-$30$\% driven by differences in the injection spectrum and oscillations.

\section{Results}
\label{sec:Results}

In this section we discuss the differential neutrino flux from WIMP annihilations in the Sun and compute event rates at detectors. We assume a generic detector and quote results per kiloton fiducial volume. We do not consider events 
interacting outside the detector volume, which have been studied extensively in previous works~\cite{Kamionkowski:1994dp,Wikstrom:2009kw,Desai:2004pq,Abbasi:2009uz}, and focus instead on vertex contained events.
The usage of vertex contained events for analyses has been well established and is documented for 
Super-Kamiokande~\cite{Ashie:2005ik}. 
Rates for these events are easily scalable to any detector size and results are to first order independent of the actual detector geometry. Rates are given as differential rates as function of energy and would have to be convoluted with the energy response and resolution for a specific detector. We chose a coarse energy binning to mimic a limited energy resolution. 
To put results in context, we compare event rates with those from atmospheric neutrinos~\cite{Gaisser:2002jj}.
 We use the angle-averaged flux and neglect atmospheric neutrino oscillations as these are most relevant at an energy range, which is largely below our region of interest and in addition results would strongly dependent on the detector location. The atmospheric tau neutrino flux component, which is almost entirely generated through neutrino oscillations, has a very strong dependence on the underlying baseline length and therefore zenith angle. 
We compare signal muon and electron neutrino rates to the corresponding component from the atmospheric neutrinos. As tau events are in general more similar in topology to electron neutrino events, we compare the tau neutrino signal rates to those from atmospheric electron neutrinos.
We assume the atmospheric neutrino background is well determined, which is a reasonable assumption as it can be estimated from the data itself from off--source regions. Further the size of the available off--source region is large compared to the on--source region, which minimizes the uncertainty on the background prediction. We compare the sensitivity of different search strategies for a generic neutrino detector assuming an angle-averaged atmospheric neutrino flux.
 We discuss other Sun related backgrounds in section~\ref{Discussion}. 


\subsection{Expected Event Rates}

We compute event rates for charged-current (CC) interactions, as they allow flavor identification and measurement of the neutrino energy. We do not consider the impact of neutral-current (NC) interactions in this work.
Deep inelastic neutrino and anti-neutrino cross sections are treated separately~\cite{Gandhi:1998ri} and we use the 
ratios of $\sigma^{\rm CC}_{\nu_{\tau}{\rm N}}/\sigma^{\rm CC}_{\nu_{\mu}{\rm N}}$ to treat 
tau neutrinos~\cite{Jeong:2010nt,PhysRevD.65.033002}.
 
We order results by neutrino flavor and compare the rates to those produced by the corresponding angular averaged atmospheric neutrino component. The differential event rate at the detector 
for each channel is be given by
\begin{equation}
\frac{dN^{\rm sig}_{\nu_{x}}}{dE} = T n \cdot  \frac{\Gamma_{\rm A}}{4\pi D_\odot ^2} \sigma_{\nu_{x} N}^{\rm CC}(E) n_{\nu_{x}}^{f}(E)  \epsilon_{\nu}, 
\end{equation}
where $n$ is the number of nucleons per kton given by $N_{\rm A}\times {\rm kton} \approx 6.023 \times 10^{32}$, where $N_{\rm A}$ is the Avogadro number. $T$ is the detector livetime in units of kton$\cdot$years. $n_{\nu_{x}}^{f}$ is the neutrino spectrum at Earth per annihilation computed with DarkSUSY as defined by equation~(\ref{eqn:diff_flux_at_detector}). 
As annihilations occur almost entirely near the center of the Sun, the resulting neutrino flux can be considered a point-like source. 
The acceptance, $\epsilon_{\nu}$, defines the fraction of events that will be contained within an opening 
angle around the Sun.

We assume three different options for the opening angle around the Sun and do not optimize them to keep results from being exposure-dependent. We use two fixed angle cuts of $30^{\circ}$ and $10^{\circ}$ and a neutrino energy dependent selection criteria $\Psi(E)$, which is used as our benchmark cut.
The opening angle contains 68\% of the signal under the assumption of the kinematic 
angle between neutrino and lepton convoluted with an intrinsic detector resolution of one degree. Hence, $\Psi(E_{\nu}) \simeq 1^{\circ} \sqrt{1 + 1{\rm TeV}/E_{\nu}}$.
This case requires that the energy of vertex contained events can be reasonably well measured on an event by event basis to extract spectral information.

\begin{figure}[htb]
\includegraphics[width=.48\textwidth]{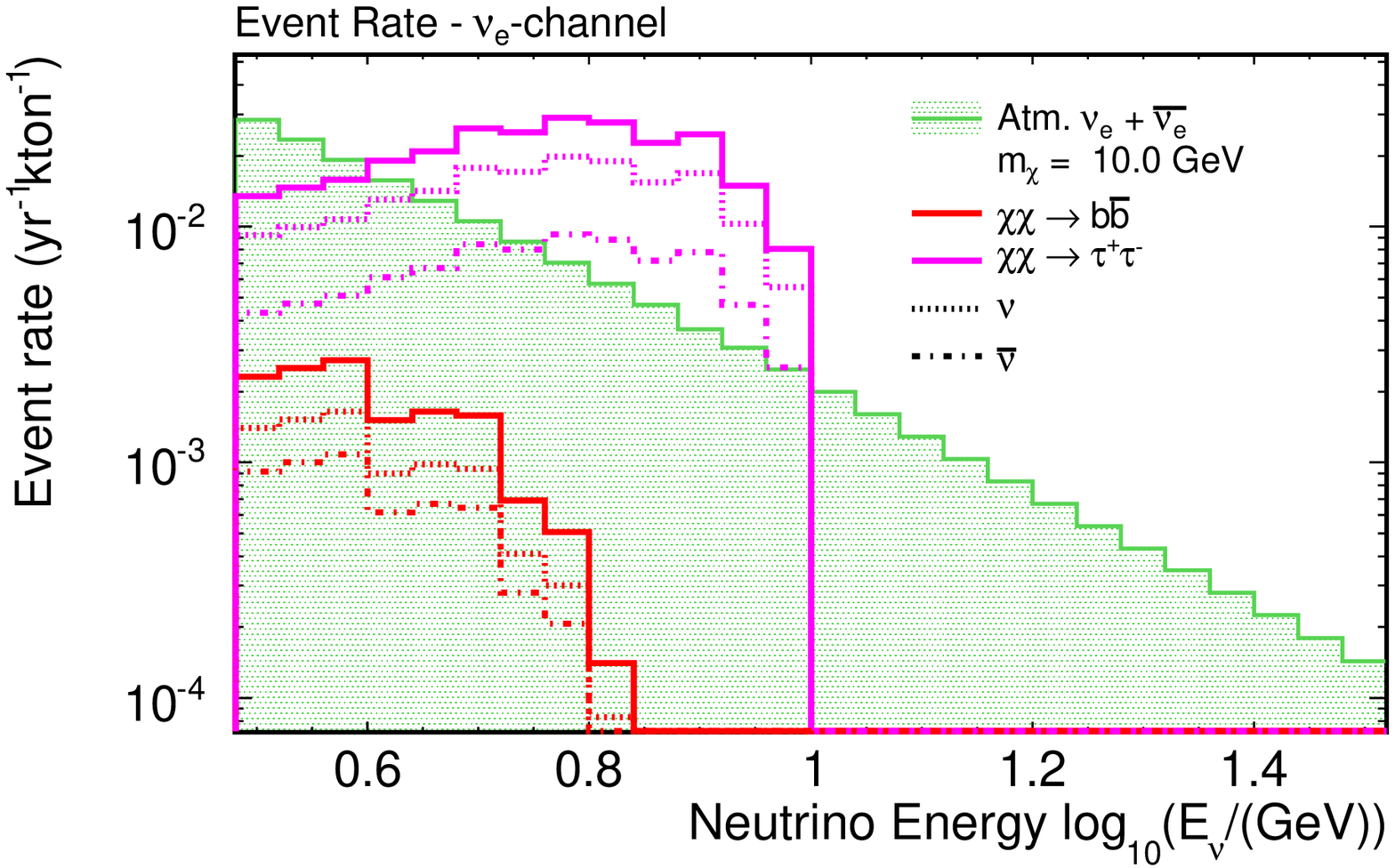}
\includegraphics[width=.48\textwidth]{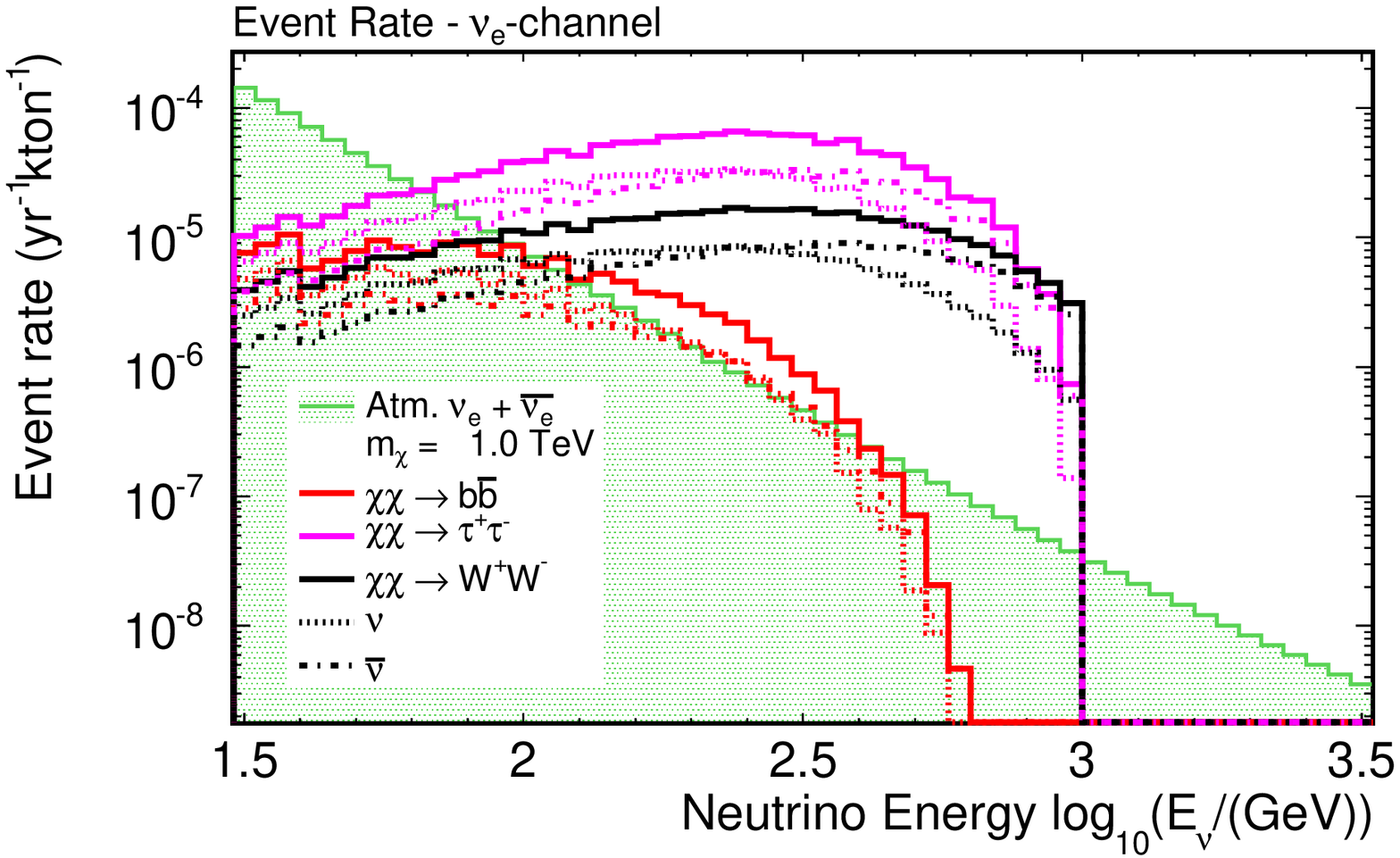}
\caption{\label{fig2}
Rates for neutrino interactions ($\nu_{\rm e} + \bar{\nu_{\rm e}}$), $\nu_{\rm e}$, and $\bar{\nu_{\rm e}}$ per kton$\cdot$years for a solar WIMP signal for various annihilation channels compared to those from atmospheric neutrinos. We assume an opening angle selection criteria containing $68\%$ of the signal. The number of events per year is given for each energy bin. The neutrino channel for $m_{\chi}=10$~GeV (left) and $m_{\chi}=1$~TeV (right) are shown for $\nu_{e}$. A spin-dependent WIMP-nucleon scattering cross section of $1.0 \times 10^{-3}$~pb~(1~fb) was assumed, which leads to an annihilation rate of $9.42\times 10^{24} {\rm s}^{-1}$, and $2.38\times 10^{21} {\rm s}^{-1}$ at equilibrium for Wimp masses of 10~GeV and 1~TeV, respectively.}
\end{figure}

In figure~\ref{fig2} we compare the differential neutrino event rates to those caused by atmospheric neutrinos. 
We use a WIMP-nucleon scattering cross section of $1.0 \times 10^{-3}$~pb, which is of the same order as present exclusion limits~\cite{Abbasi:2009uz,Desai:2004pq}. For the corresponding annihilation rate please refer to table~\ref{equi_table}. 
We show the rates as function of neutrino energy for neutrinos and anti-neutrinos separately and compare results with the atmospheric neutrino background.
The number of background events depends on the angular opening angle around the Sun; hence the background is given by:
\begin{equation}
\frac{dN^{\rm bkg}}{dE}=T n \sigma_{N \nu_{x}}^{CC}(E) \phi_{\nu_{x}}^{\rm atm}(E) \int_0^{\Psi(E)} 2\pi \cos(\theta) d\theta.
\end{equation}

For cascade-like events the rate would therefore correspond roughly to the number of fully contained events, while for the muon neutrino flux the given rate corresponds to the fully and partially contained events. We have cross-checked our estimates against reported multi-GeV event rates of the corresponding types at Super-Kamiokande~\cite{Wendell:2010md} and found good agreements.

\begin{figure}[htb]
\begin{minipage}[t]{.49\textwidth}
    \includegraphics[width=.99\textwidth]{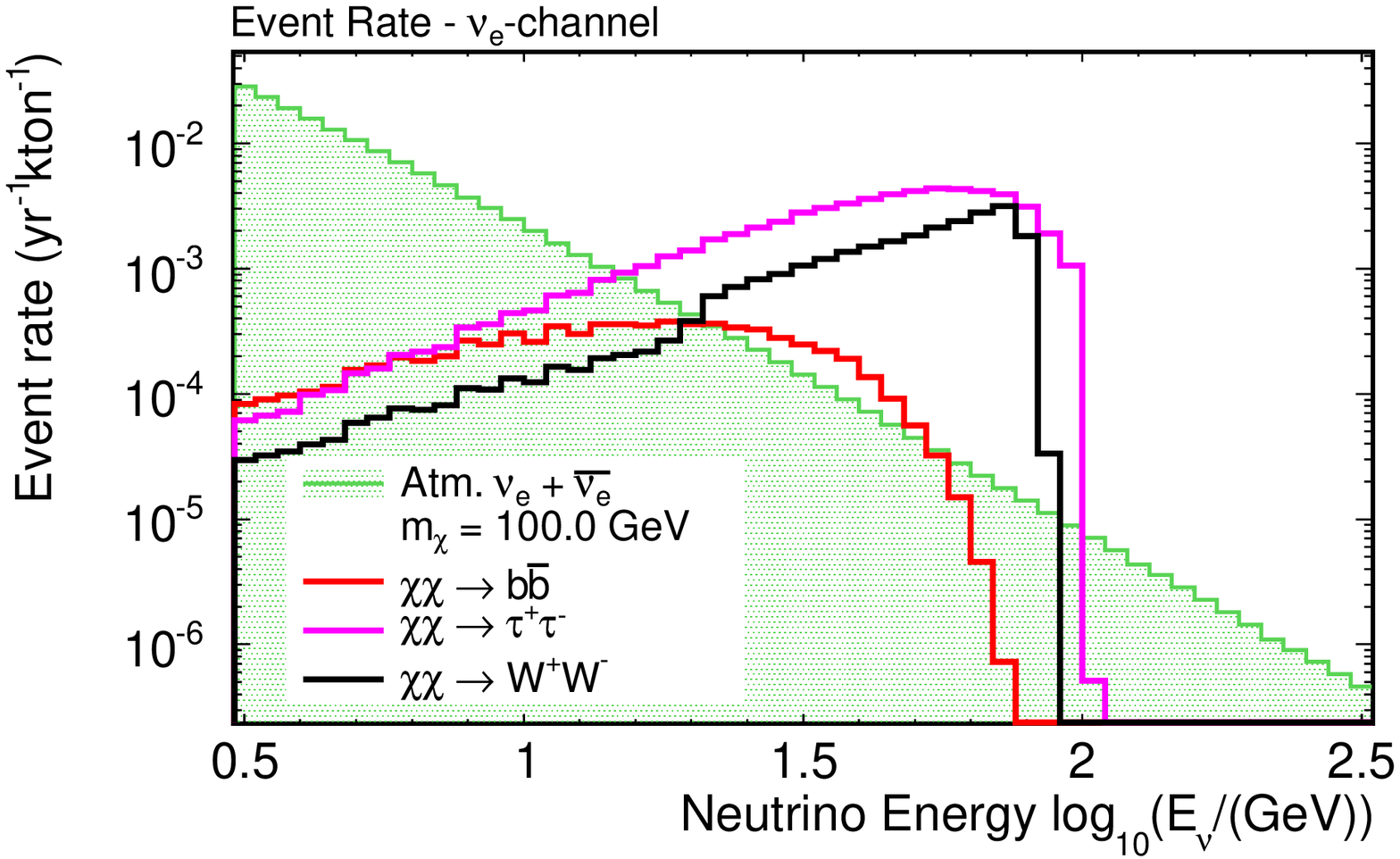}
  \end{minipage}
  \hfill
\begin{minipage}[t]{.49\textwidth}
    \includegraphics[width=.99\textwidth]{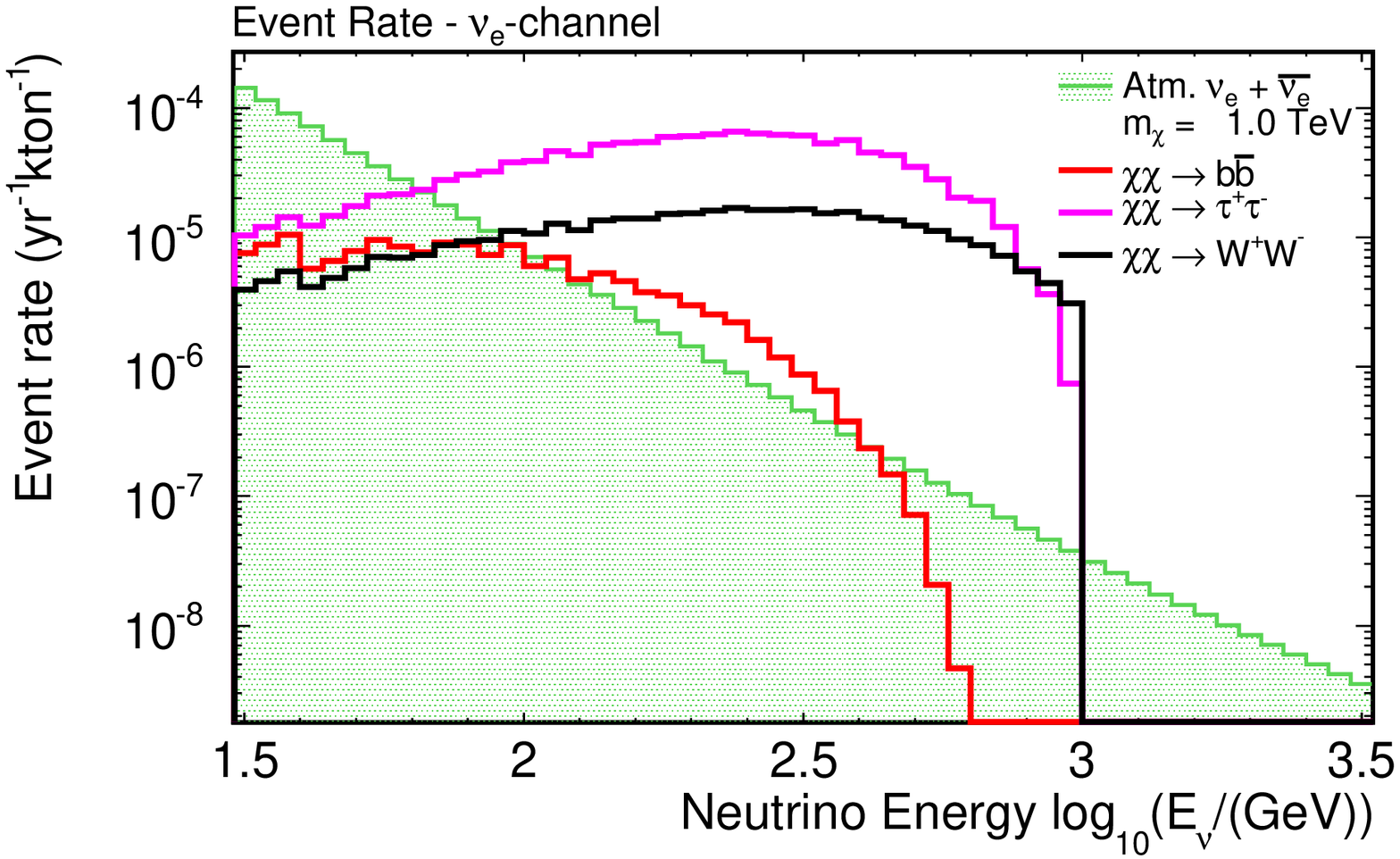}
  \end{minipage}
\begin{minipage}[t]{.49\textwidth}
    \includegraphics[width=.99\textwidth]{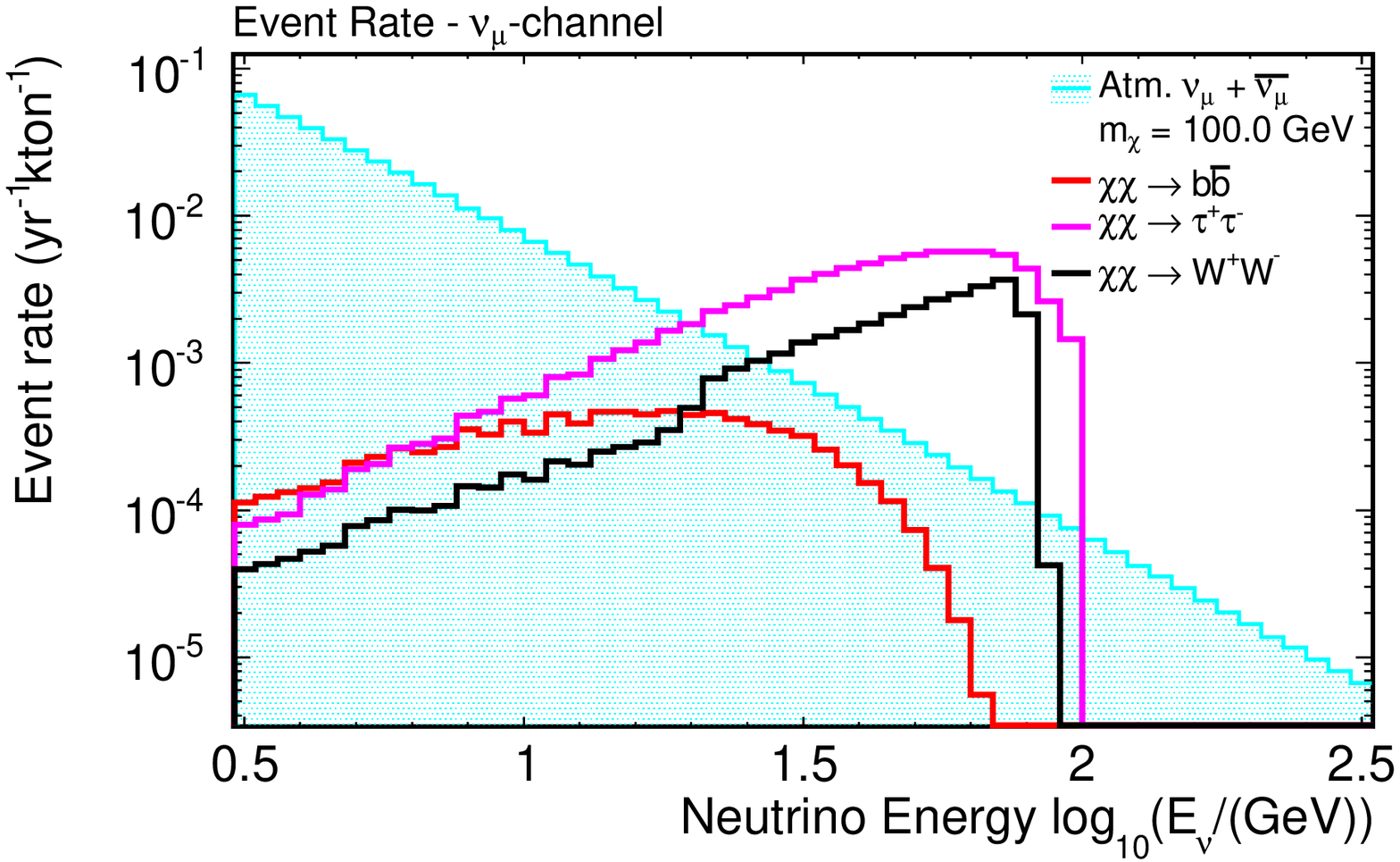}
  \end{minipage}
  \hfill
\begin{minipage}[t]{.49\textwidth}
    \includegraphics[width=.99\textwidth]{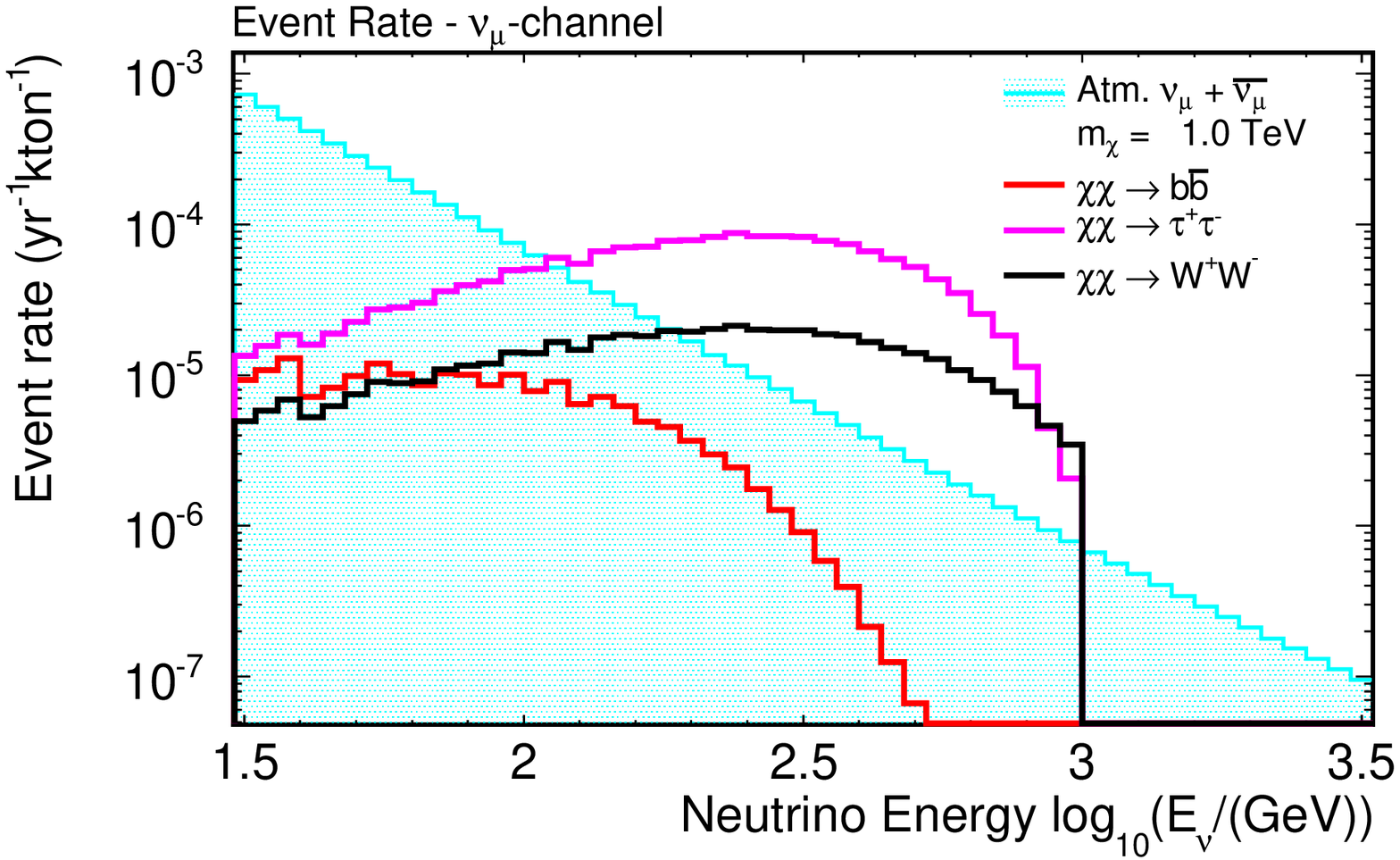}
  \end{minipage}
\begin{minipage}[t]{.49\textwidth}
    \includegraphics[width=.99\textwidth]{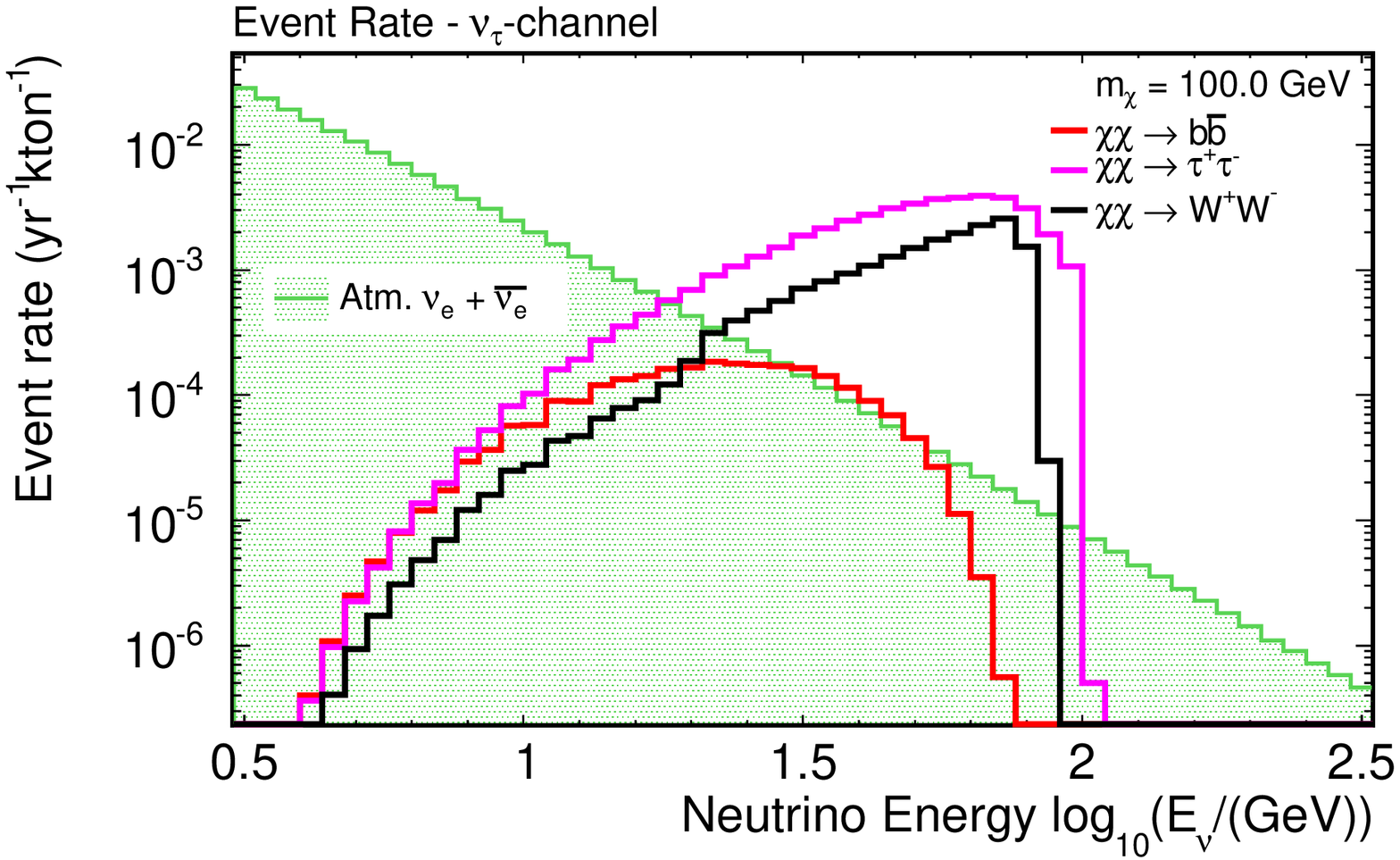}
  \end{minipage}
  \hfill
\begin{minipage}[t]{.49\textwidth}
    \includegraphics[width=.99\textwidth]{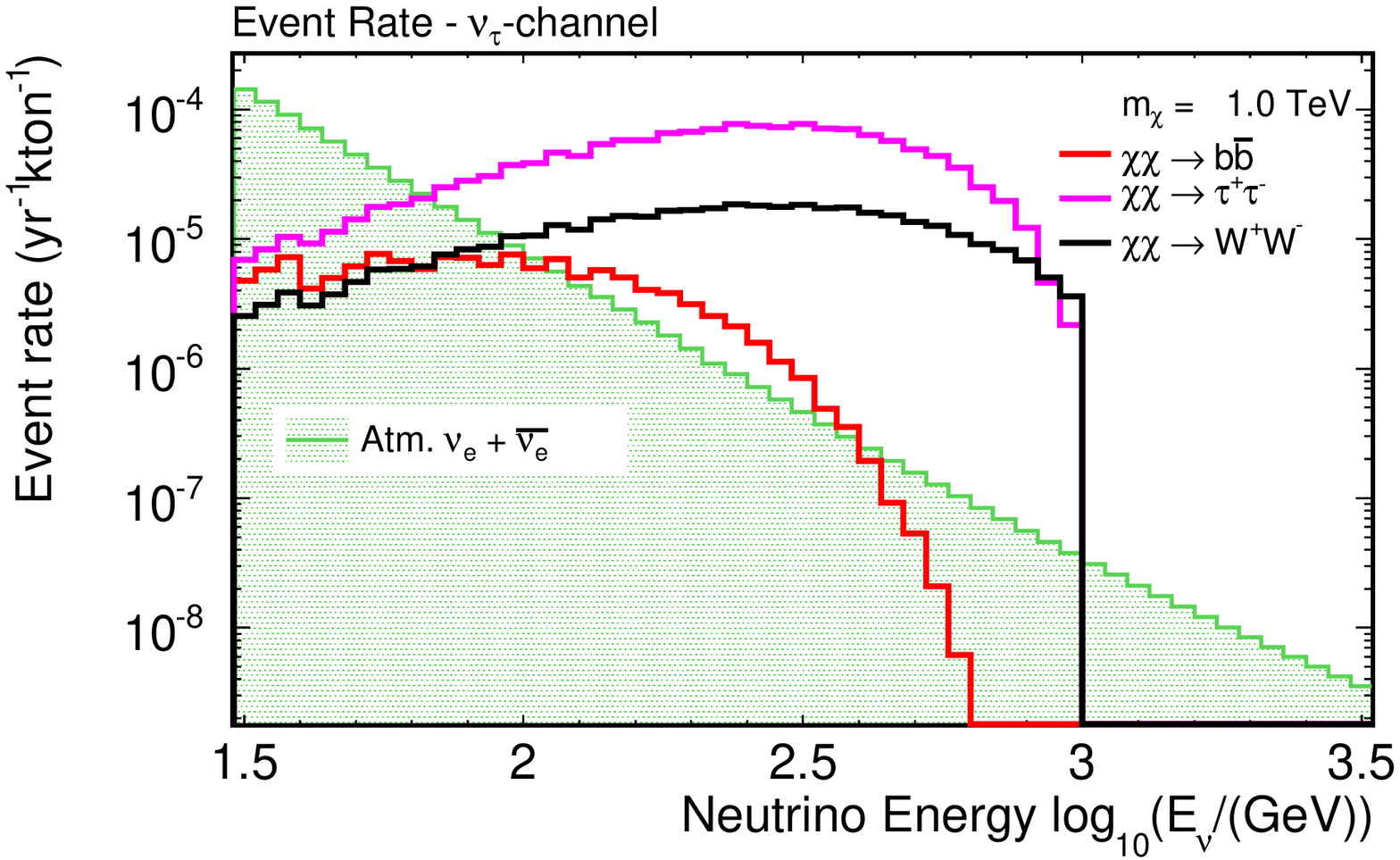}
  \end{minipage}
\caption{\label{fig3}
Rates per kton$\cdot$years for neutrino interactions ($\nu + \bar{\nu}$) from atmospheric neutrinos are compared to those  from a solar WIMP signal for various annihilation channels. Neutrinos are selected within an opening angle around that Sun contains $68\%$ of the WIMP signal. The event rates are given per energy bin for the three active neutrino flavor channels $\nu_{e}$, $\nu_{\mu}$, and $\nu_{\tau}$ (top to bottom) and for two different WIMP masses $m_{\chi}=100$~GeV(left) and $m_{\chi}=1$~TeV (right). A spin-dependent WIMP-nucleon scattering cross section of $1.0 \times 10^{-3}$~pb~(1~fb) was assumed, which leads to a capture rate of $2.45 \times 10^{23}~1/{\rm s}$ ($m_{\chi}=100$~GeV) and $2.38 \times 10^{21}~1/{\rm s}$ ($m_{\chi}=1$~TeV).}
\end{figure}

Since neutrino detectors are generally not sensitive to charge, except magnetized designs like MINOS~\cite{Michael:2008bc}, we give rates for the sum of neutrinos and anti-neutrinos. Figure~\ref{fig3} shows event rates per kton$\cdot$year for CC interactions by neutrino flavor compared to the corresponding background, which would result in track or cascade-like events. 
As an example we have calculated event rates for 200~kton$\cdot$years, which roughly corresponds to the total integrated lifetime of
14~years of Super-Kamiokande data, see table~\ref{table_2} for event rates assuming a lower energy threshold of 1~GeV. 
Hyper-K or similar next generation water Cherenkov detectors would acquire a comparable lifetime in a year or less~\cite{Suzuki:2001rb,Nakamura:2003hk}.
Rates need to be compared with about 300 (646) atmospheric electron (muon) neutrino events for a threshold energy of 1~GeV. For energies above 10~GeV rates drop to 2 (8) for electron (muon) neutrinos.

\begin{table}[h]
\caption{\label{table_2}Integrated number of events in 200~kton$\cdot$years, assuming a half cone opening angle containing 68\% of the signal. Using a lower energy threshold of 1~GeV we have integrated $\nu_{e},\nu_{\mu},\nu_{\tau}$ over an energy range of 1~GeV to 10~TeV. The total rate is broken down in the that from neutrinos and anti-neutrinos separately in the bracket. A spin-dependent WIMP-nucleon scattering cross section of $1.0 \times 10^{-3}$~pb was assumed, which leads to a capture rate of $2.45 \times 10^{23}~1/{\rm s}$ ($m_{\chi}=100$~GeV) and $2.38 \times 10^{21}~1/{\rm s}$ ($m_{\chi}=1$~TeV).}
\begin{center}
\begin{tabular}{|l|r|r|r|r|}
\hline
Flavor & WIMP mass & \multicolumn{3}{|c|}{Events per 200~kton$\cdot$years} \\
channel & (GeV) & $\chi\chi\rightarrow b\bar{b}$ & $\chi\chi\rightarrow \tau^{+}\tau^{-}$ & $\chi\chi\rightarrow {\rm W}^{+}{\rm W}^{-}$\\
\hline
$\nu_{e}$ &      10   & 9.1   (5.5/3.5) & 67.1 (45.8/21.3)     & 0.0 (0.0/0.0)  \\ 
 &              100   & 1.5   (0.9/0.6) & 12.4  (7.9/ 4.5)     & 5.3 (3.3/2.0)  \\ 
 &             1000   & 0.05 (0.03/0.02) &  0.28  (0.15/ 0.13) & 0.08 (0.04/0.04)  \\
\hline
$\nu_{\mu}$ &    10   & 12.5 (8.8/3.7) & 86.8 (58.3/28.5) & 0.0 (0.0/0.0)  \\ 
 &              100   & 1.9 (1.3/0.6) & 16.4 (10.4/ 6.0) & 6.6 (4.2/2.4)  \\ 
 &             1000   & 0.05 (0.03/0.02) &  0.37  (0.19/ 0.18) & 0.10 (0.05/0.05)  \\
\hline
$\nu_{\tau}$ &   10   & 0.02 (0.01/0.01) &  2.5  (1.8/ 0.7) & 0.0 (0.0/0.0)  \\ 
 &              100   & 0.50 (0.34/0.16) &  9.0  (5.8/ 3.2) & 3.8 (2.4/1.4)  \\ 
 &             1000   & 0.03 (0.02/0.01) &  0.30  (0.15/ 0.15) & 0.08 (0.04/0.04)  \\
\hline
\end{tabular}
\end{center}
\end{table}

\subsection{Sensitivity}

We compute the sensitivity for a generic neutrino detector assuming a data sample, of 200~kton$\cdot$years and $5$~Mton$\cdot$years, as could be achieved by a future detector such as Hyper-K~\cite{Nakamura:2003hk}, LENA~\cite{Wurm:2011zn}, or an infill array to IceCube/DeepCore. For the sensitivity study we compare the expected neutrino spectrum from background only with that containing in addition to the background a signal. We assume that the WIMP mass and dominant annihilation channel is unknown, and hence we consider the entire energy spectrum. In general one can assume that a signal would be clustered around a few adjacent energy bins. A sliding energy window would in this case be an optimal search strategy, but it inherently also generates trial factors that need to be considered. As contributions to the background are tiny from high energy events due to the steeply falling atmospheric neutrino spectrum, we consider an energy threshold rather than a window. 
For simplicity we use three different energy thresholds: 1~GeV for $m_{\chi}\le 100$~GeV,
10~GeV for $1~{\rm TeV} > m_{\chi}\ge 100$~GeV, and 100~GeV threshold
for $m_{\chi}\ge 1$~TeV. As we expect significant absorption 
in the Sun at energies close to 1~TeV, moving the threshold any higher is not desired.

\begin{figure}[htb]
\includegraphics[width=.48\textwidth,angle=-90]{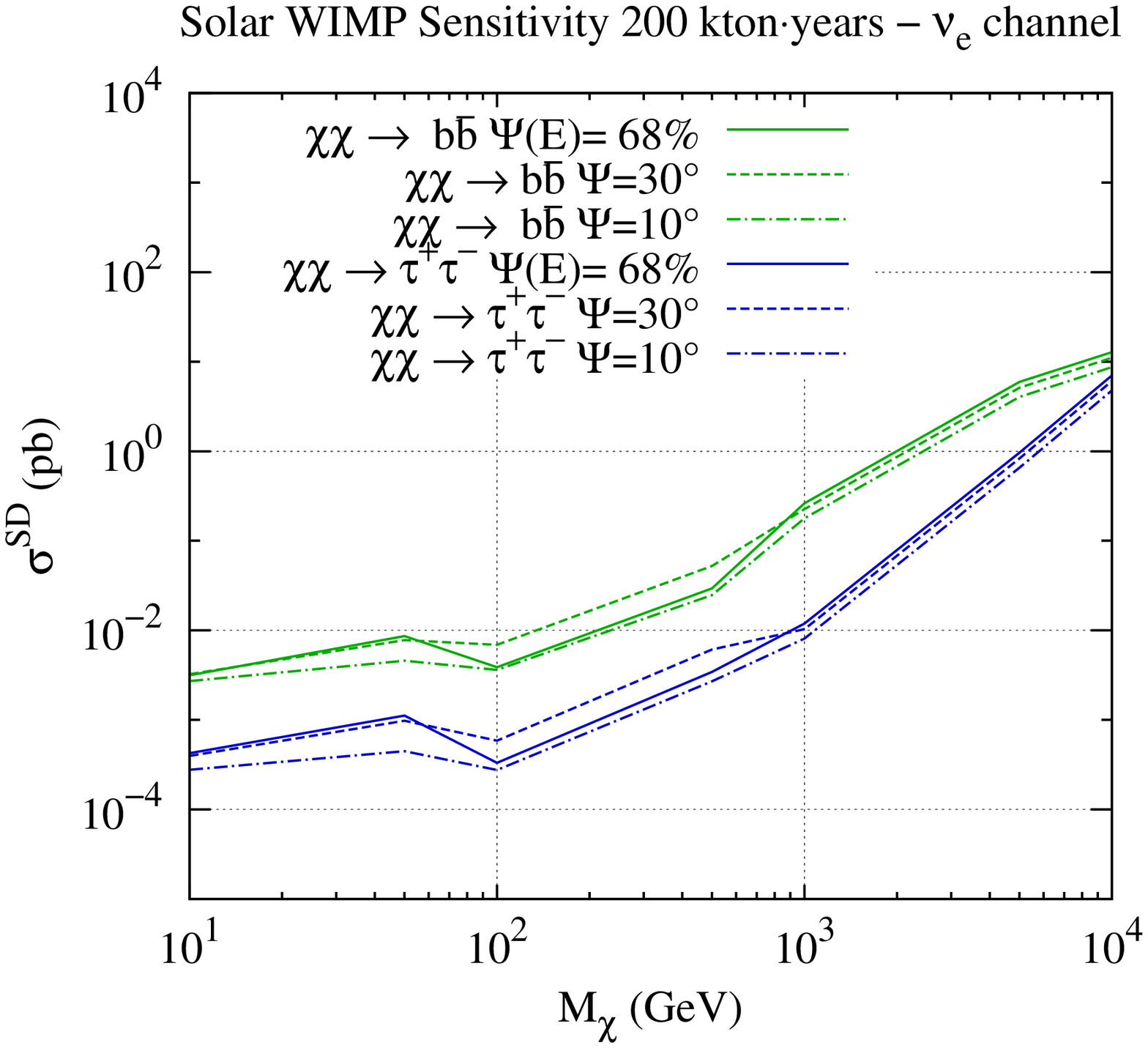}
\includegraphics[width=.48\textwidth,angle=-90]{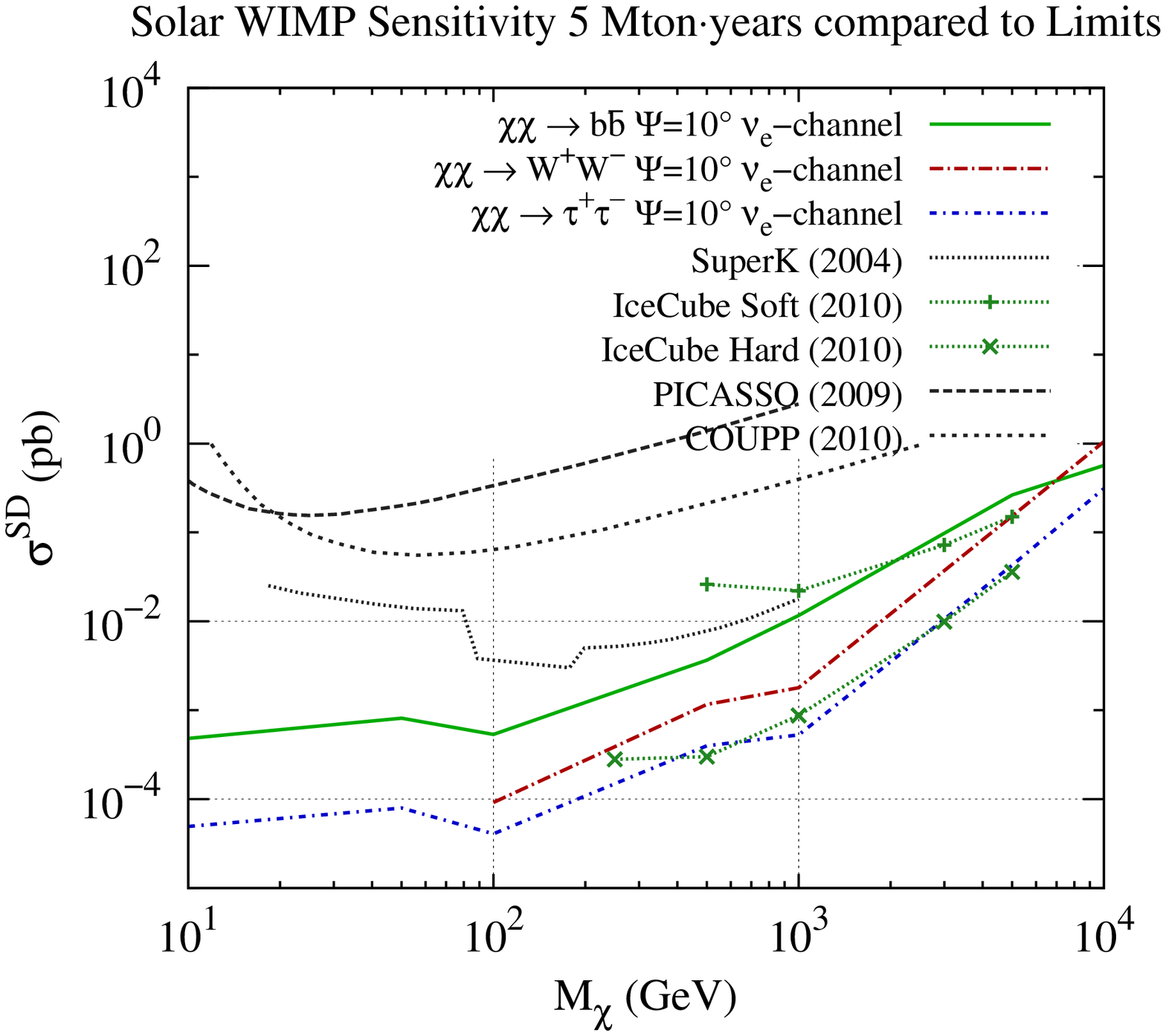}
\includegraphics[width=.48\textwidth,angle=-90]{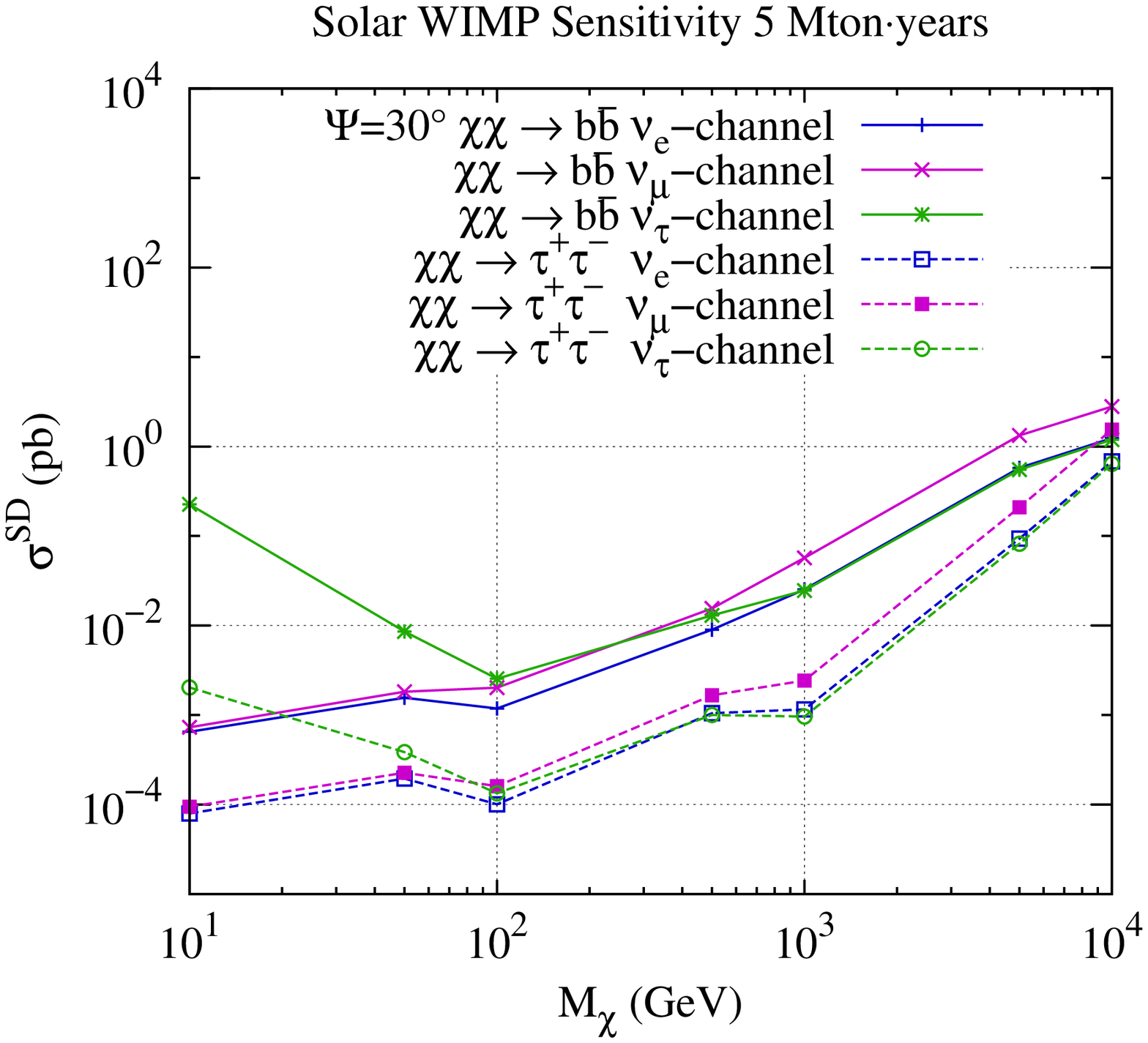}
\includegraphics[width=.48\textwidth,angle=-90]{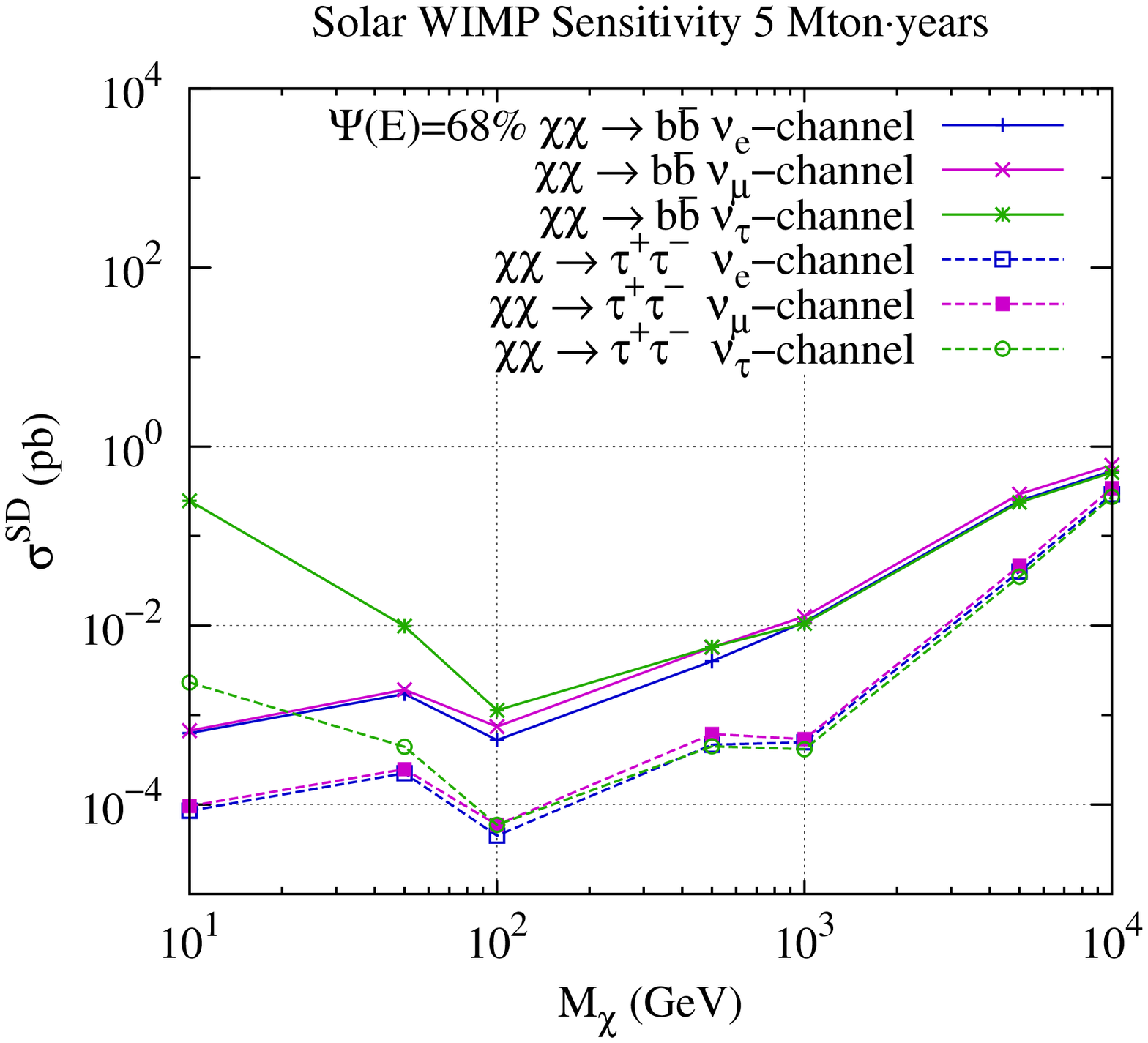}
\caption{\label{fig4}
Top left: Sensitivity for a data size of 200~kton$\cdot$years for annihilations into $\tau^{+}\tau^{-}$ and $b\bar{b}$. Three different analysis methods using a fixed $30^{\circ}$ cone centered on the Sun, $10^{\circ}$ cone, and an adaptive cone that contains 68\% of the signal for each energy bin. Top right: Sensitivity for 5~Mton$\cdot$years for annihilations into $\tau^{+}\tau^{-}$, $W^{+}W^{-}$, and $b\bar{b}$ compared to present bounds from direct and indirect detection experiments. Limits from Super-Kamiokande (SuperK)~\cite{Desai:2004pq}, IceCube~\cite{Abbasi:2009uz}, COUPP~\cite{Behnke:2010xt}, and PICASSO~\cite{Archambault:2009sm} are shown. Bottom: Comparison of the different neutrino flavor channels for 5~Mton$\cdot$years for an opening angle of $30^{\circ}$ (left) and one containing 68\% of the signal (right).
}
\end{figure}

We determine the sensitivity based on the average 90\%C.L. Feldman-Cousins~\cite{Feldman:1997qc} upper limit that would be obtained by an ensemble of experiments with no signal in the presence of a known background~\cite{Hill:2002nv}. For large statistics (number of background events above 100), we approximate the expected limit with Gaussian statistics via $\sqrt{b}\times 1.645$. The assumption that the background is well known is valid for our case as it can be determined from the data itself using an off--source region which is significantly larger than the target region.
As the size of the signal depends directly on the WIMP-nucleon scattering cross section, we determine the corresponding $\sigma^{\rm SD}$ for the limit on the expected event rate.
The obtained sensitivities are shown in figure~\ref{fig4}. We compare three different methods: A fixed $30^{\circ}$ cone centered on the Sun, a fixed $10^{\circ}$ cone, and an adaptive cone that contains 68\% of the signal for each energy bin as described above. Using a small or adaptive cone size generally performs best.

Assuming a comparable angular resolution is achieved for all neutrino flavors, the electron neutrino channel generally outperforms other flavor channels. The tau neutrino channel is also very compelling for WIMP masses above 100~GeV, while at lower WIMP masses rates are significantly decreased due to the smaller tau neutrino cross section~\cite{Jeong:2010nt,PhysRevD.65.033002}. Tau events can nevertheless be important as they could be detected nearly background free based on their unique topology. We have neglected NC induced events, which will further increase signal rates. 

\subsection{Discussions}
\label{Discussion}

Besides the atmospheric neutrino background, Solar WIMP searches are also prone to irreducible neutrino backgrounds from the Sun itself. Interactions of cosmic-rays with the solar atmosphere have been predicted to
produce $\gamma$-rays and high energy neutrinos~\cite{Seckel:1990pc,Ingelman:1996mj,Moskalenko:2006ta}.
These albedo neutrinos are produced through the decay of charged pions in the solar atmosphere. Flux predictions differ widely and are beyond the scope of this paper. Hence, we only give the magnitude of this flux, which is estimated to be about 0.01-0.04~events per year per Mton for each neutrino flavor assuming an energy threshold of 10~GeV~\cite{Fogli:2006jk}. These rates are however still small compared to the corresponding signal rates for the achieved sensitivities.
Transient backgrounds caused by solar flare neutrinos, which can reach energies up to a few 10~GeVs with a peak flux at around 100~MeV for a few minutes, generally fall below the relevant neutrino energies discussed in this work, but might need to be considered in analyses~\cite{Hirata:1988sx}.

Previous searches have focused on the muon neutrino induced muon flux at neutrino telescopes for obvious reasons: (1) the good pointing resolution of high-energy muons allows us to define a small search window around the Sun. It can further be used in rejecting the large background caused by down-going atmospheric muons. (2) The long range of high-energy muons (for example $\sim 1$~km at 200~GeV in water/ice), allows to detect a flux of muons caused by $\nu_{\mu}$ interactions outside the instrumented detector volume. 
Advances in detection methods with operating neutrino detectors and prospects for future detectors make it feasible to expand Solar WIMP searches to other neutrino flavors. The case of using electron neutrinos in Liquid Argon detectors have been discussed recently~\cite{Kumar:2011hi}, and obtained a similar sensitivity compared to results presented here. Not only could these provide additional information on the WIMP properties, in case of a detected signal, but also searches relying on vertex contained events can generally achieve a better energy resolution.
Further, if the muon range is of the same order as the size of the detector, then there is no longer a significant increase in effective area due its range. 
We also note that neutrinos fluxes with energies above 1~TeV are severely attenuated and neutrinos above 100~GeV encounter significant absorption effects in the Sun. Hence, neutrino searches are limited to an energy range below a few hundred GeV.

\section{Discussions of Sources of Uncertainty}
\label{sec:Uncertainty}

In this section we discuss uncertainties on the conversion of the neutrino
flux to the WIMP-nucleon scattering cross section and provide a description for the treatment of them. 

We group the systematic uncertainties into three different categories that are uncorrelated to each other:
(1) uncertainty on the annihilation rate at equilibrium, $\Gamma_{\rm A}^{\rm equi}$,
(2) uncertainty on the neutrino flux related to the neutrino propagation to Earth,
(3) uncertainty related to the flux measurement at the detector and propagation through Earth. 

Through this grouping, the error treatment can be simplified as categories are independent of each other and can be considered separately. In addition different experimental results using this description can easily be compared directly.
While, the first two categories are common to all neutrino detectors, the third category is highly detector dependent. We focus our discussions on the first two categories and merely provide a summary of the relevant effects for the third category. A detailed treatment would have to be performed by the corresponding experiments.
 
For the uncertainty discussion we assume the equilibrium condition and address the impact of systems 
that are perturbed from it separately. This is part of a discussion on the equilibration time scale,
WIMP-nucleon scattering cross section and self-annihilation cross section in section~\ref{impact_of_non_equilibrium}.


\subsection{Annihilation Rate Uncertainty}

At equilibrium the annihilation rate is solely determined by the capture rate, $\Gamma_{\rm A}^{\rm equi} = \frac{1}{2} \Gamma_{\rm C}$.
Hence, we can use the uncertainty on $\Gamma_{\rm A}$ at equilibrium to group all uncertainties related to the capture and annihilation process.
A variety of uncertainties need to be considered, that are related to the dark matter halo density and velocity distribution at the solar Galactic orbit, capture mechanism, and composition of the Sun. We summarize the uncertainties here and quantify them where possible.

First of all, the capture rate in the Sun depends linearly on the local dark matter density.
Hence, under the assumption of a constant capture rate as resulting from a smooth dark matter halo, the uncertainty on 
$\Gamma_{\rm A}$ is directly dependent on the uncertainty due to the halo density:
$\frac{\Delta \Gamma_{\rm A}}{\Gamma_{\rm A}} \simeq \frac{\Delta  \rho_{0}}{\rho_{0}}$. Note that this uncertainty affects direct detection experiments in the same way.
In our studies we have adopted a local dark matter halo density of $\rho_{0}=0.3$~GeV/cm$^3$ at the location of the Sun, which is commonly used~\cite{Nakamura:2010zzi}. 
However, recent measurements favor a higher value of $0.39\pm0.03$~GeV/cm$^3$~\cite{Catena:2009mf}. Simulations of Milky Way type galaxies, that include baryons, leads to a significant flattening of the dark matter halo in the direction normal to the stellar disk. Pato et al.~\cite{Pato:2010yq} found that this effect could lead to a dark matter overdensity in the local disk of up to 41\% and estimate an average enhancement of 21\% for such a triaxial profile compared to a spherically symmetric Einasto profile, resulting in $\rho_{0} = 0.466 \pm 0.033({\rm stat}) \pm 0.077({\rm syst}) {\rm GeV}/{\rm cm}^3$.
Based on various methods a general be bound can be found for $0.25~{\rm GeV/cm^3}< \rho_{0} < 0.70~{\rm GeV/cm^3}$~\cite{Salucci:2010qr}.

\begin{figure}[htb]
\includegraphics[angle=-90,width=.48\textwidth]{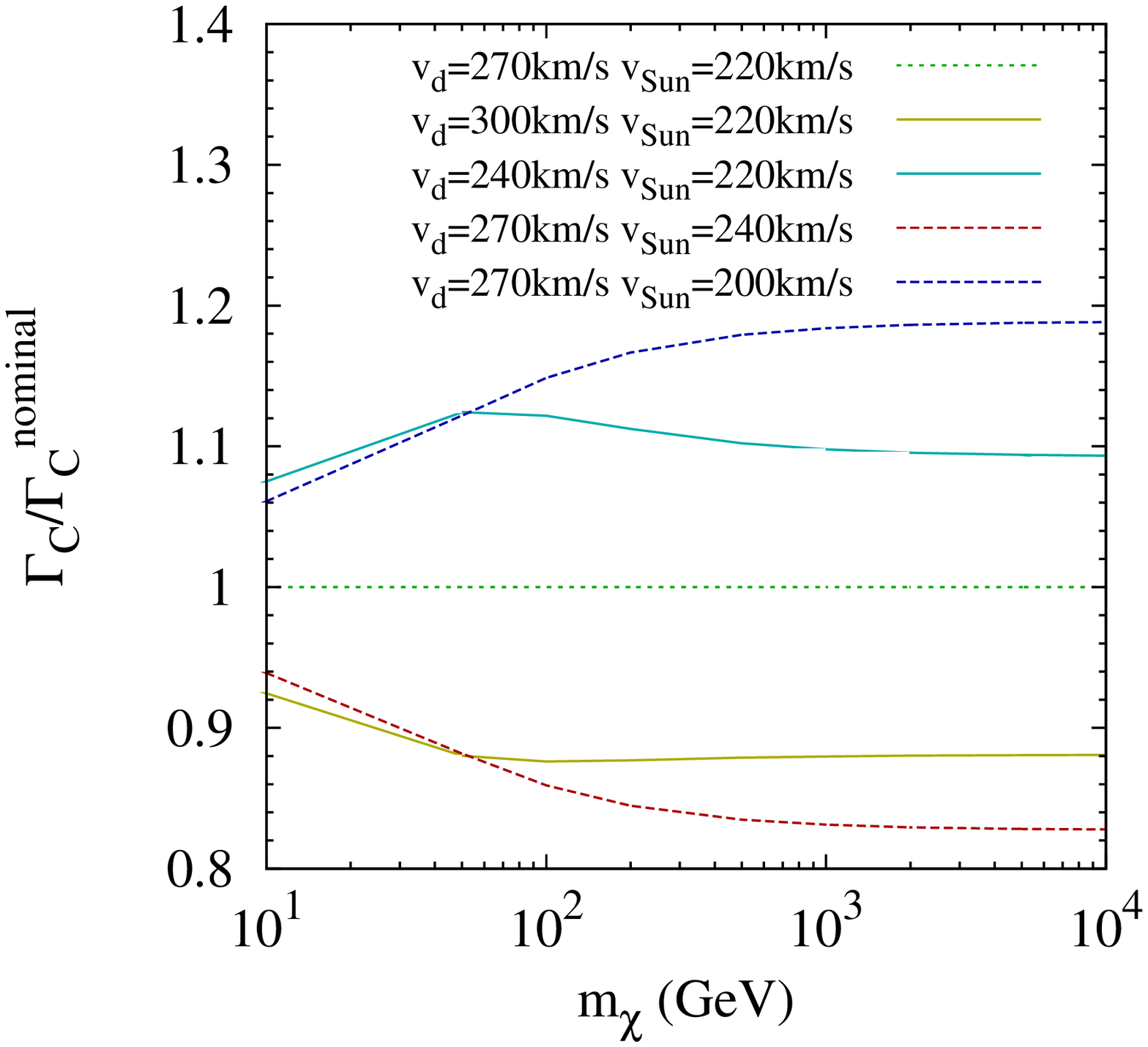}
\includegraphics[angle=-90,width=.48\textwidth]{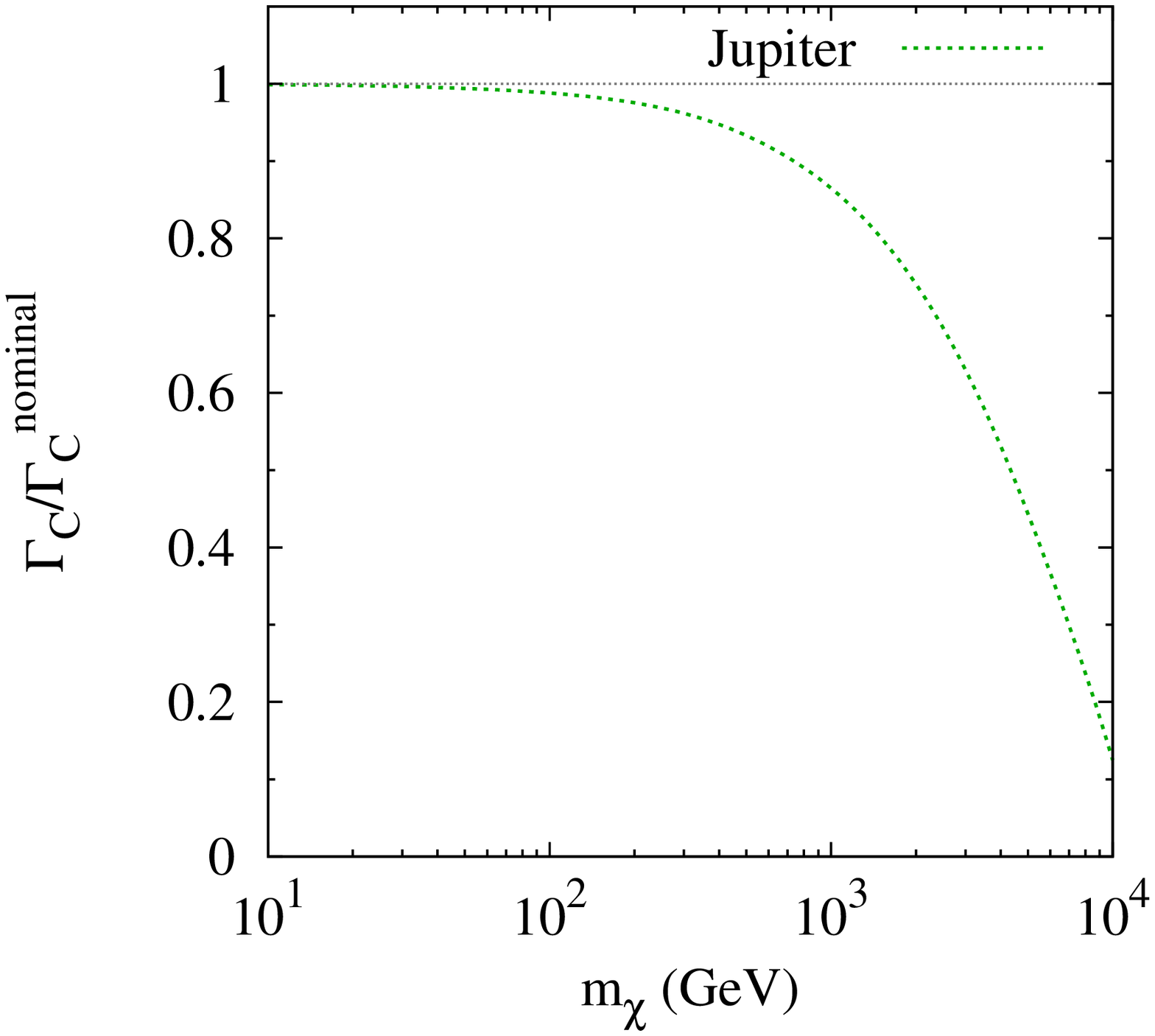}
\caption{
Left: Change in capture rate as function of WIMP mass and halo parameters compared to the nominal model assuming a circular velocity of 220~km/s and a velocity dispersion of 270~km/s.
Right: Impact of Jupiter on the capture rate as function of the WIMP mass.
\label{fig5}}
\end{figure}

The capture rate depends in a complex manner on the circular velocity of the Sun (relative to the halo) and the 
WIMP velocity dispersion, which is commonly assumed to be Maxwellian as given by equation~(\ref{Maxwell_vel}). We find that the uncertainties related to the velocity dispersion and circular velocity is decreased for smaller WIMP masses.  Figure~\ref{fig5} shows the change in capture rate for various halo parameters differing from our default Maxwellian velocity distribution with a dispersion of $v_{d}=270$~km/s and velocity of the Sun relative to the halo is 
assumed to be $v_\odot=220$~km/s. Note, that recent simulations~\cite{Vogelsberger:2008qb} indicate deviations from Maxwellian velocity distributions. 
For a discussion of the impact of different velocity distributions on the capture rate, we refer the reader to Bruch et al.~\cite{Bruch:2009rp}, which have studied this effect in light of the merger history of the Milky Way. 
Implications for direct-detection experiments that are mostly sensitive to the high velocity tail have been discussed~\cite{Lisanti:2010qx}. Contrarily for the capture rate in the Sun the low velocity region is most relevant. 
It was found that the annihilation rate in the Sun could be enhanced by as much as an order of magnitude relative to standard halo models if the ratio of the accreted dark matter is of the same order as the halo dark matter density. This increase owes to the higher phase space density at low velocities in the accreted dark disc. 

Galactic substructure, or subhalos, would naturally lead to varying dark matter density and velocity distributions that can be address with the same error treatment described above, assuming extremes between dense and depleted regions. The assumption that the capture rate is constant with time is no longer true for this case and hence the annihilation rate would also be perturbed from its nominal value. Extreme cases of substructure and the impact on capture rates and equilibration time scale has been discussed elsewhere~\cite{Koushiappas:2009ee}. Contributions from additional dark matter populations (for example in form of a dark disk~\cite{Bruch:2009rp}) as the Milky Way could have acquired during its complex merger history falls in the same category and is not discussed in further detail. 
Deviations from a smooth halo profile at the solar circle through substructure would introduce a time dependence on the strength of the neutrino flux from the Sun. N-body simulations disfavor any significant sub structure on scales that can currently be resolved at the solar circle~\cite{Vogelsberger:2008qb}. There are indications for a more complex structure in the velocity distribution. For the case of the Sun small variations in the local dark matter density and velocity distribution can generally be neglected compared to the overall uncertainty on the average dark matter density and velocity distribution, as the expected annihilation rate in the Sun is better defined by the dark matter capture integrated over a time scale similar to the equilibrium time scale.  

The dynamics of the solar system impact the capture process and can result in a change in the capture rate~\cite{Peter:2009mk}. Modeling of this effect is extremely complex and has only been attempted for a simplified three-body system, taking the gravitational effects of Jupiter into account, which is expected to be the dominant effect. Based on these studies suppression in the capture rate was observed. 
Taking additional planets into account in simulations will further reduce the capture rates, though Jupiter is expected to have the largest impact~\cite{private_peter}.
The capture rate suppression shows a strong dependence on the WIMP mass and becomes relevant for WIMP masses in the TeV range as shown in figure~\ref{fig5}. To take this reduction into account one can define a WIMP mass dependent suppression factor
$\alpha({\rm m}_{\chi})=(\Gamma_{\rm C}/\Gamma_{\rm C}^{\rm nominal})^{-1}$. The uncertainty on
this suppression factor directly propagates to the uncertainty in $\Gamma_{\rm A}$. Due to the complexity of the evaluation of planetary effect the uncertainty is not well understood. Conservatively one could set an upper bound on the uncertainty by comparing the capture rates with and without the effect from Jupiter taken into account. While this might be practical for WIMP masses below one TeV, where the difference is less than $15\%$, a more detailed study would be needed for higher masses. 

Uncertainties in the composition of the Sun will also alter the capture rate, as well as a possible spin-independent component. However, the uncertainty in the elemental abundances in the Sun only affects the flux predictions on the percent level~\cite{Ellis:2009ka}.

Table~\ref{table_of_uncertainites} summarizes the systematic uncertainties on the annihilation rate. We consider three different cases as examples, a low WIMP mass 10~GeV, a 100~GeV WIMP, and a high WIMP mass 1~TeV. Uncertainties are given with respect to the assumed default parameters: $\rho = 0.3$~GeV/cm$^{3}$, and a Maxwellian velocity distribution with a velocity dispersion of 270~km/s and a solar circular velocity of 220~km/s. While, some uncertainties can be large, they generally result in an increased annihilation rate, so that constraints tend to be conservative. The two dominant effects, the velocity distribution and local dark matter density affects direct detection experiments at the same magnitude. While, the local dark matter density affects rates at direct detection experiments in the same way, the impact of the velocity distribution can have very different affects. 

\begin{table}[h]
\caption{\label{table_of_uncertainites} This table summarizes the systematic uncertainties on the annihilation rate $\Delta \Gamma_{\rm A}/ \Gamma_{\rm A}$ at equilibrium with respect to the used default values and their uncertainty. For details see text and references. This summary table is meant to give a rough idea on the uncertainties and their approximate size, an analysis would require a more detailed discussion of them including considering correlations between some of the quantities.
}
\begin{center}
\begin{tabular}{|l|r|r|r|r|}
\hline
Uncertainty & \multicolumn{3}{|c|}{WIMP Mass $m_{\chi}$} & Dependence \\
            &  $10$~GeV & $100$~GeV & $1$~TeV & / Source \\
\hline
Dark matter density &  ${ }^{+130\%}_{-17\%}$ & ${ }^{+130\%}_{-17\%}$ & ${ }^{+130\%}_{-17\%}$ & $\Delta \Gamma_{\rm A} /\Gamma_{\rm A} \approx \Delta \rho_{0} / \rho_{0}$ \cite{Salucci:2010qr,Catena:2009mf,Pato:2010yq}\\
Capture process (Planets) & $<1\%$ &$\sim 1$\% & $\pm 20$\% & (see Fig~\ref{fig5} (right)) \cite{Peter:2009mk,Peter:2008sy,Gondolo:2004sc} \\
Solar composition         &  $\sim 1$\% & $\sim 1$\% & $\sim 1$\% & \cite{Ellis:2009ka} \\
Solar velocity            &  $\pm 6$\% & $\pm 15$\% & $\pm 18$\%  & (see Fig~\ref{fig5} (left)), \cite{Gondolo:2004sc} \\
Velocity dispersion       & $\pm 8$\% & $\pm 12$\%&  $\pm 10$\%  &(see Fig~\ref{fig5} (left)),  \cite{Gondolo:2004sc} \\
Velocity distribution     & \multicolumn{3}{|c|}{Large enhancements possible}  & \cite{Bruch:2009rp} \\
Evaporation               & small & $\sim 0$\% &  $\sim 0$\%  & \cite{Gaisser:1986ha,Griest:1986yu}\\
\hline
\hline
\end{tabular}
\end{center}
\end{table}

\begin{figure}[htb]
\includegraphics[width=.48\textwidth]{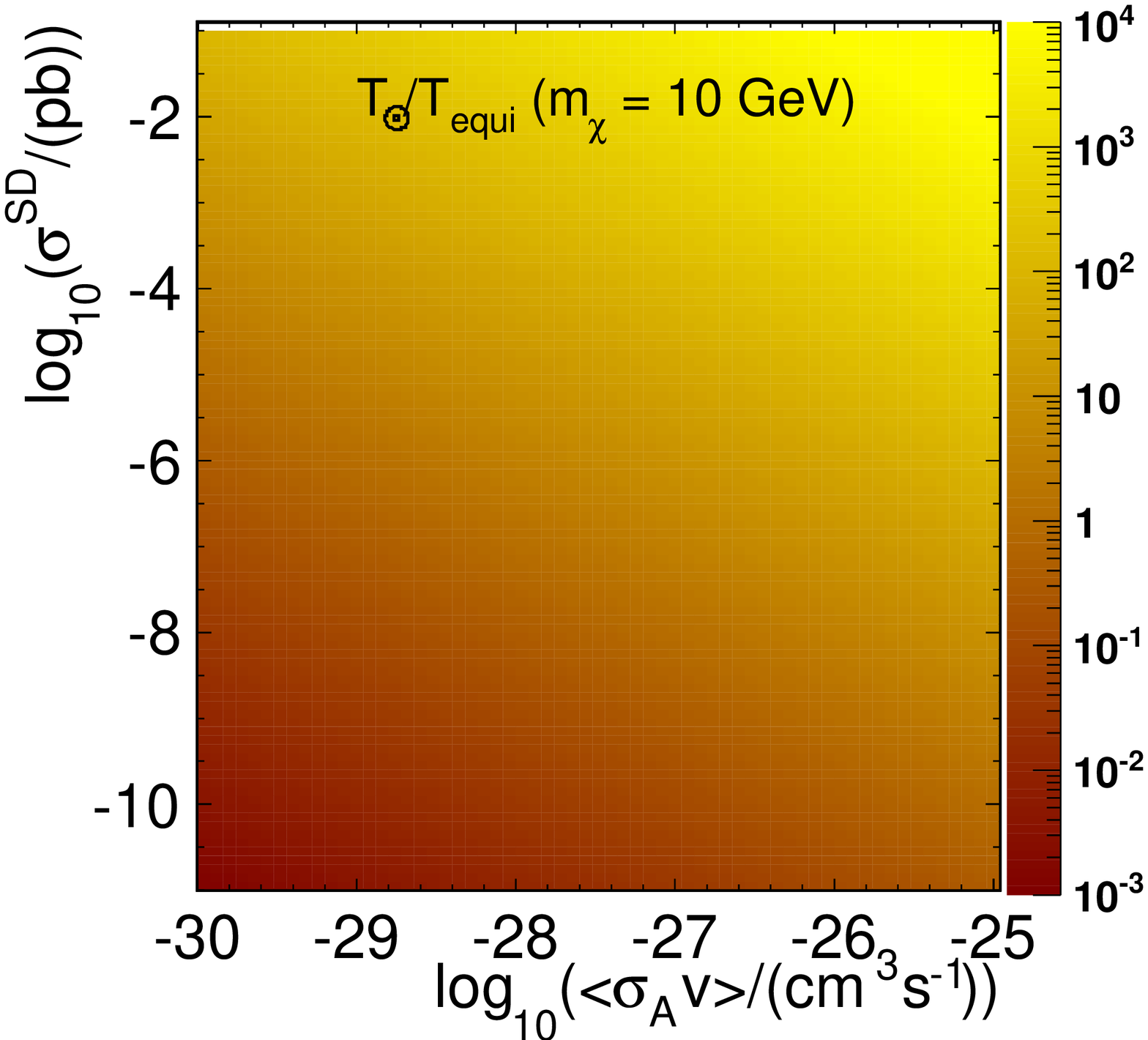}
\includegraphics[width=.48\textwidth]{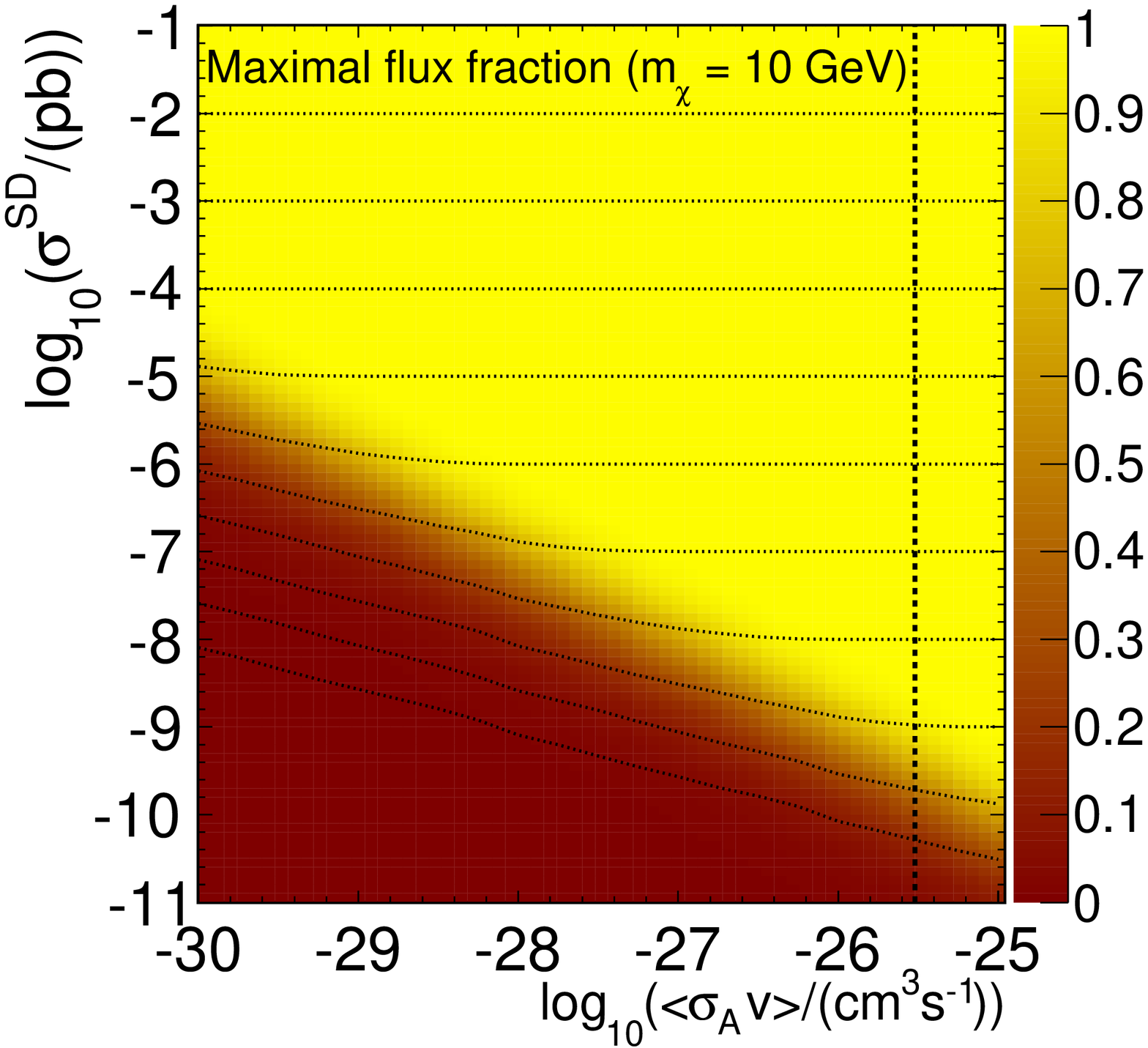}
\caption{Left: Average factor by which the age of the Sun exceeds the equilibration time scale, as function of the self-annihilation
  cross section and the WIMP-nucleon spin-dependent scattering for a WIMP of mass 10~GeV. Right: The corresponding fraction of the neutrino flux compared to the maximal flux reached at equilibrium. The dotted line indicates the thermal relic cross section, while the dashed lines indicate the affect on a constrained on the WIMP Nucleon cross section if the dark matter self-annihilation cross section is reduced so that the Sun is not in equilibrium anymore.
\label{fig6}}
\end{figure}

\label{impact_of_non_equilibrium}
While the equilibration time scale is typically short compared to the age of the Sun, in some scenarios the Sun might not be equilibrium. In this case the annihilation rate would be reduced to a fraction of the maximal rate achieved at equilibrium. The expected neutrino flux is reduced by the same factor. We determine the ratio of the expected flux compared to the maximal flux.
The average flux fraction obtained via DarkSUSY is shown in figure~\ref{fig6}. It can be seen that the equilibrium condition is in general satisfied for the region of parameter space accessible by current and planned experiments.

For a detailed discussion on astrophysical uncertainties we would also like to refer the reader to Serpico and Bertone~\cite{Serpico:2010ae}.

\subsection{Neutrino Propagation Uncertainty}

The second class of uncertainties, which can be treated independently from the uncertainty on the annihilation rate, $\Gamma_{\rm A}$, is related to the propagation of neutrinos from the center of the Sun to Earth. 

The uncertainties on the neutrino flux, $\phi_{\nu}^{f}$, are related to neutrino oscillation and absorption effects as well as regeneration of (tau) neutrinos in the Sun. These effects are fully implemented in DarkSUSY. 
For a general discussion on the neutrino fluxes~\cite{Cirelli:2005gh} from dark matter annihilation and their flavor ratios~\cite{Lehnert:2007fv} we refer the reader to the associated references.

For neutrino production at the center of the Sun, absorption is relevant for neutrinos above 100~GeV, and leads to a significant suppression for neutrinos above 1~TeV. 
The effects of neutrino oscillations and interactions have been studied elsewhere~\cite{Blennow:2007tw} using an event-based full three flavor Monte Carlo. It was concluded that for neutrinos the main effects of neutrino oscillations are the effective mixing of muon and tau neutrinos during the propagation to the solar surface and the consequent mixing of electron neutrinos during the propagation from the surface of the Sun to the Earth. The mixing between electron and other neutrino flavors is not complete, due to the leptonic mixing angle $\theta_{12}$ being less than maximal, which results in the yields for muon and tau neutrinos to be similar but differ from those of the electron flavor. The impact of the oscillation and propagation effects can vary based on the annihilation channels and WIMP masses, and hence need to be considered for individual cases. As an example the systematic uncertainty due to neutrino oscillation and propagation effects was found to be $4\%$ for the IceCube Solar WIMP analysis~\cite{Abbasi:2009uz,Wikstrom:2009zz}. Note also that the energy resolution of a typical neutrino telescope is limited, which leads to an effective averaging of any fast oscillation effects.

\subsection{Detector related Systematic Uncertainties}

Solar WIMP searches hold the advantage that background rates can often directly be taken from an off--source region, which reduces systematic uncertainties associated with the background estimate significantly. Evaluating the signal acceptance is however often a more complicated undertaking. Since the impact of detector related uncertainties highly dependent on the experiment, we can just provide an overview of the most relevant effects. 

Underlying to all detectors is the uncertainty on the neutrino interaction cross section, which has uncertainties of a few percent in the GeV - TeV energy region~\cite{Pumplin:2002vw}.
Somewhat larger is the uncertainty for $\nu_{\tau}$($\bar{\nu_{\tau}}$) that is known to about $6\%$($8\%$)~\cite{Jeong:2010nt} at the 10~GeV range.  
Neutrino propagation effects through the Earth introduce another source of uncertainty.

The uncertainty on the energy measurement is another contribution for analyses that rely on measurements of the neutrino energy.
Considering a heavily instrumented water Cherenkov experiments energy resolutions that have been achieved are $\pm3\%$ for a single muon ring and $\pm(2.5/\sqrt{E/{\rm GeV}} +0.5)\%$ for a single electron
ring~\cite{Shiozawa:1998si}.
Further for large water/ice Cherenkov detectors also detector specific systematic effects are often dominated by photon propagation in the detector medium, 
limited knowledge of the acceptance and absolute efficiency of the individual photomultiplier tubes, as well as reconstruction related uncertainties~\cite{Desai:2004pq,Abbasi:2009uz}.
For vertex contained events, however systematic uncertainties have been reduced to about $5\%$~\cite{Wendell:2010md}.


\section{Conclusions}
\label{sec:Conclusions}

We have studied the comparison of neutrino fluxes from the Sun with the spin-dependent WIMP-nucleon cross section using DarkSUSY. The use of conversion factors for neutrino fluxes compared to muon fluxes hold some distinct advantages.
Inherently, they do not depend on muon propagation effects and allow us to treat all neutrino flavors in the same way. 
We have discussed a procedure for the treatment of systematic uncertainties associated with the flux conversion and investigated the impact of these uncertainties. While, associated uncertainties on the annihilation rate can be significant they are not limiting to measurements.

Using vertex contained events and spectral event information we have estimated the sensitivity for neutrino telescope with a corresponding datasets of 200~kton$\cdot$years and 5~Mton$\cdot$years. For WIMP masses below 100~GeV the sensitivity to the spin-dependent WIMP-nucleon scattering cross section extends below $10^{-4}$~pb. Vertex contained events out perform muon rates once the detector size is on the same order as the corresponding track length of the muon neutrino induced muon. Even for higher WIMP masses, competitive results can be achieved using relatively large opening angles around the Sun that could be used to compensate for limited angular resolution.
Assuming a comparable angular resolution is achieved the electron neutrino channel generally outperforms other flavor channels.


\begin{acknowledgments}
We would like to thank John Beacom, Joakim Edsj{\"o}, and Annika Peter for many useful discussions and Basu Dasgupta, Choe Goun, Shunsaku Horiuchi, and Matthew Kistler for comments on the paper.
This work was supported through the CCAPP visitor program and by the GCOE Program QFPU of Nagoya University from JSPS and MEXT of Japan. 
\end{acknowledgments}

\newpage

\bibliographystyle{JHEP}
\bibliography{JCAP_Rott_Tanaka_Itow_ref}

\end{document}